\documentclass[journal,onecolumn]{IEEEtran}
\IEEEoverridecommandlockouts
\usepackage[utf8]{inputenc}
\usepackage{amsthm}
\usepackage{amsfonts,amssymb,latexsym,cite}
\usepackage[cmex10]{amsmath}
\usepackage{stmaryrd}
\usepackage{algorithmic}
\usepackage{array}
\usepackage{bbm}
\usepackage{verbatim}
\usepackage{color,xcolor}
\usepackage{graphicx}
\usepackage{float}
\usepackage{epstopdf}
\usepackage[font={small,it}]{caption}
\usepackage[T1]{fontenc}
\usepackage{url}
\usepackage{enumerate}
\usepackage{enumitem}
\usepackage{epsfig}
\usepackage{soul}
\usepackage{thmtools,thm-restate}
\usepackage[margin=1in]{geometry}
\usepackage{tikz}
\usetikzlibrary{arrows,arrows.meta,automata,shapes,calc,intersections,fit,positioning}
\usepackage{pgfplots}
\pgfplotsset{compat=newest}
\usepackage[hidelinks]{hyperref}
\hypersetup{
  pdftitle={Exact Second-Order Asymptotics in Covert Communication Over Discrete Memoryless Channels},
  pdfsubject={Variational distance and missed detection at a fixed false-alarm probability}
}
\newtheorem{definition}{Definition}

\allowdisplaybreaks
\begin{document}
\title{Exact Second-Order Asymptotics for Covert Communication over DMCs with Variational Distance Constraints}
\author{Qiaosheng Zhang\thanks{Qiaosheng Zhang is with the Shanghai AI Laboratory and Shanghai Innovation Institute.}, 
\and Lin Zhou, \emph{Member, IEEE}\thanks{Lin Zhou is with the Southern University of Science and Technology.}, and \and Xuelong Li, \emph{Fellow, IEEE}\thanks{Xuelong Li is with the Institute of Artificial Intelligence (TeleAI) of China Telecom.} }
\date{}
\maketitle

\begin{abstract}
We determine the exact second-order asymptotics of covert communication over binary-input discrete memoryless channels when covertness is measured by variational distance. Previous work by Tahmasbi and Bloch
[IEEE Trans. Inf. Theory, Apr. 2019] characterized the first-order
asymptotics and derived achievability and converse bounds on the
second-order term, but these bounds do not match. The gap arises from
an additional penalty of order \(n^{1/4}\) in the achievability bound. We show that this penalty can be removed through a sharper analysis of the distribution of the warden's output induced by pulse-position modulation. Specifically, we express the variational distance through the Bhattacharyya coefficient of two distributions and the expectation of a continuous function of the log-likelihood ratio. Because the resulting expectation involves a continuous function rather than the probability of a likelihood-ratio event, an analysis of the characteristic function combined with a Gaussian smoothing argument reduces the approximation error from \(O(n^{-1/4})\) (derived from the Berry--Esseen bound in prior work) to  \(O(n^{-1/2})\). With this better controlled approximation error, we manage to derive a matching achievability result to the existing converse result, thus establishing the exact second-order asymptotics.

\end{abstract}

\begin{IEEEkeywords}
Covert communication, second-order asymptotics,  variational distance, pulse-position modulation.
\end{IEEEkeywords}

\setcounter{page}{1}

\section{Introduction}
Covert communication considers the situation in which a transmitter communicates reliably with a legitimate receiver while concealing the presence of communication from a warden. Since the square-root law was formalized for additive white Gaussian noise channels in~\cite{Bash2013}, the information-theoretic limits of covert communication have been progressively characterized for binary symmetric channels~\cite{Che2013, ZhangBakshiJaggi2021}, discrete memoryless channels (DMCs)~\cite{Bloch2016,Wang2016}, classical-quantum channels~\cite{WangCQ2016,Sheikholeslami2016}, multiple-access channels~\cite{Arumugam2016,Arumugam2019MAC}, broadcast channels~\cite{Arumugam2019Broadcast, TanLee2018, SteinbergWigger2026}, channels with state~\cite{Lee2018}, adversarial channels~\cite{ZhangBakshiJaggi2019}, MIMO AWGN channels~\cite{WangBloch2021,LiuWangZhangXuZhou2026}  and continuous-time channels subject to spectral constraints~\cite{Zhang2019}. These studies show, under the usual non-degeneracy conditions, that only \(O(\!\sqrt n\,)\) information bits can be communicated reliably and covertly over \(n\) channel uses, and also identify the exact pre-constant before \(\sqrt n\), which is commonly referred to as the \emph{covert capacity}. For the binary-input DMC considered here, pulse-position modulation (PPM) provides a structured signaling mechanism that attains the optimal first-order scaling~\cite{BlochGuha2017}.

The finite blocklength behavior is less completely understood. For binary-input DMCs, Tahmasbi and Bloch \cite{Tahmasbi2019} developed a unified second-order analysis under maximum probability of error for three covertness metrics: KL divergence, variational distance, and the probability of missed detection at a fixed probability of false alarm. Their achievability part, for all the three metrics, relies on a random coding argument with the PPM scheme~\cite{BlochGuha2017}. When covertness is measured by KL divergence, their achievability and converse bounds match up to the second term in the order of $n^{1/4}$, yielding exact  second-order asymptotics. For the other two measures including variation distance, the authors of  \cite{Tahmasbi2019} identified the exact first-order term, of order \(n^{1/2}\), but the corresponding achievability and converse bounds differ for the second-order term scaling in the order of $n^{1/4}$. Consequently, the exact second-order asymptotics remained open for these two covertness metrics.

In particular, variational distance has been widely adopted as a covertness metric in the literature, including in pioneering works such as \cite{Bash2013,Che2013}, because it has a direct operational interpretation in terms of the false-alarm and missed-detection probabilities of the warden’s hypothesis test. A standard binary hypothesis testing argument shows that, when the variational distance is small, the sum of the false-alarm and missed-detection probabilities remains close to one even when the warden Willie uses optimal detectors.

The objective of this work is to establish the exact second-order asymptotics for covert communication under the variational distance metric. We show that the achievable second-order term attained by pulse-position modulation (PPM)~\cite{BlochGuha2017} coincides with the converse term established in \cite{Tahmasbi2019}, hence deriving the exact second-order asymptotics. The coding scheme remains unchanged from that work; the improvement lies entirely in a sharper analysis of the output distributions induced at the warden when the legitimate transmitter transmits or not. The proof of \cite[Lemma 8]{Tahmasbi2019} applies the Berry--Esseen theorem separately to two likelihood-ratio events, resulting in an approximation error of order \(\ell^{-1/2}\), where \(\ell\) denotes the number of PPM pulses. Since \(\ell=\Theta(\sqrt n)\), this error is of order \(n^{-1/4}\). To ensure that the covertness constraint is met despite an error of this scale,  the number of pulses must be reduced by at least \(\Theta(n^{1/4})\). Since the leading term of the logarithm of the achievable message size is proportional to \(\ell\), reducing \(\Theta(n^{1/4})\) pulses reduces the logarithm of the message size by \(\Theta(n^{1/4})\), which is exactly the order of the second-order term. Therefore, the achievability bound fails to match the converse \cite{Tahmasbi2019} at the second order.

Our analysis combines the two likelihood-ratio contributions at the level of their overlap before introducing a Gaussian approximation. We first use the elementary identity
$1-\mathbb V(P,Q)=\sum_z \min\{P(z),Q(z)\},$
and then rewrite the pointwise minimum in terms of the \emph{Bhattacharyya coefficient}. This yields a \emph{midpoint representation} of $1 - \mathbb{V}(P,Q)$  as the product of the Bhattacharyya coefficient and the expectation of the continuous function \(e^{-|x|/2}\) under a suitably defined midpoint distribution (see Lemma~1). The resulting expectation is analyzed through a characteristic-function analysis combined with Gaussian smoothing and an anti-concentration argument. These steps give an approximation error of order \(O(n^{-1/2})\). Since the Gaussian approximation depends on the normalized pulse count \(\ell/\sqrt n\), an error of order \(O(n^{-1/2})\) in the covertness expression can be compensated by changing \(\ell/\sqrt n\) by \(O(n^{-1/2})\), which corresponds to an \(O(1)\) adjustment in \(\ell\). Such an adjustment does not affect the \(n^{1/4}\) coefficient of the achievable message size. Consequently, the achievable second-order term coincides with the converse term.

The rest of this paper is organized as follows. Section II introduces the system model and states the main results. Section III establishes a sharp approximation, in variational distance, between the PPM-induced distribution of the warden's output and the innocent output distribution. Section IV combines these approximations with the coding scheme of \cite{Tahmasbi2019} to obtain the exact second-order asymptotics, and Section V concludes the paper.

\begin{figure}[H]
\centering
\resizebox{0.7\linewidth}{!}{\begin{tikzpicture}[
    x=1cm,
    y=1cm,
    >=Latex,
    line width=0.75pt,
    channel/.style={draw, minimum width=2.35cm, minimum height=1.15cm,
                    align=center, font=\large},
    terminal/.style={draw, rounded corners=2pt, minimum width=2.05cm,
                     minimum height=0.72cm, align=center,
                     font=\sffamily\normalsize},
    subsystem/.style={draw, densely dotted, rounded corners=3pt,
                      inner sep=7pt},
    signal/.style={font=\large},
    smallsignal/.style={font=\normalsize}
]

\node[terminal] (encoder) at (1.65,2.05) {ENCODER $f$};
\node[signal, anchor=east] (message) at (-0.05,2.05) {$W$};
\draw[->] (message.east) -- (encoder.west);

\coordinate (encoded) at (3.45,2.05);
\coordinate (innocent) at (3.45,0.88);
\coordinate (selector) at (4.18,1.47);
\draw (encoder.east) -- (encoded);
\fill (encoded) circle (1.5pt);
\node[signal, anchor=east] (zero) at (3.02,0.88) {$\mathbf 0$};
\draw (zero.east) -- (innocent);
\fill (innocent) circle (1.5pt);
\fill (selector) circle (1.5pt);
\draw (selector) -- (3.52,1.98);
\node[font=\scriptsize, anchor=west] at (3.53,2.13) {$H_1$};
\node[font=\scriptsize, anchor=west] at (3.53,0.73) {$H_0$};

\node[subsystem, fit=(message)(encoder)(zero)(encoded)(innocent)(selector),
      label={[font=\small]above:Transmitter}] (txbox) {};

\node[channel] (mainchannel) at (6.65,2.05) {$W_{Y|X}^{\otimes n}$};
\node[channel] (wardenchannel) at (6.65,0.18) {$W_{Z|X}^{\otimes n}$};
\coordinate (fork) at (4.88,1.47);
\draw (selector) -- (fork);
\draw[->] (fork) |- (mainchannel.west);
\draw[->] (fork) |- (wardenchannel.west);
\node[signal, anchor=south] at (4.62,1.55) {$\mathbf X$};

\node[terminal] (decoder) at (10.05,2.05) {DECODER $\varphi$};
\draw[->] (mainchannel.east) -- node[above, signal] {$\mathbf Y$} (decoder.west);
\node[signal, anchor=west] (what) at (11.72,2.05) {$\widehat W$};
\draw[->] (decoder.east) -- (what.west);
\node[subsystem, fit=(decoder)(what),
      label={[font=\small]above:Legitimate receiver}] (rxbox) {};

\coordinate (keyleft) at (1.65,3.40);
\coordinate (keyright) at (10.05,3.40);
\draw[->] (keyleft) -- (encoder.north);
\draw (keyleft) -- (keyright);
\draw[->] (keyright) -- (decoder.north);
\node[signal, fill=white, inner sep=1.5pt] at (5.85,3.40) {$S$};

\node[terminal] (detector) at (10.05,0.18) {DETECTOR};
\draw[->] (wardenchannel.east) -- node[above, signal] {$\mathbf Z$} (detector.west);
\coordinate (decision) at (11.68,0.18);
\draw[->] (detector.east) -- (decision);
\node[anchor=west, align=left, font=\normalsize] (hypotheses) at (11.82,0.18)
  {$H_0:\ Q_0^{\otimes n}$\\[2pt]
   $H_1:\ \widehat P_{Z^n}$};
\node[subsystem, fit=(detector)(hypotheses),
      label={[font=\small]above:Warden}] (wardenbox) {};

\end{tikzpicture}}
\caption{System model of covert communication over a binary-input DMC.}
\label{fig:system-model}
\end{figure}
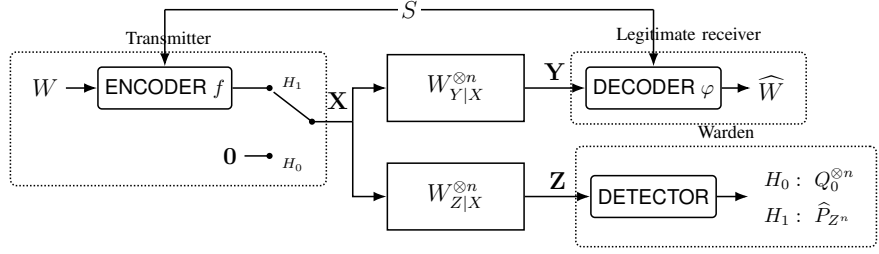

\section{Model And Main Results}
\subsection{Notation}
Random variables are denoted by upper-case letters, while their realizations are denoted  by lower-case letters. Vectors are written in boldface. All logarithms and exponentials are natural. For integers \(a\le b\), let
$\llbracket a,b\rrbracket
\triangleq
\{a,a+1,\ldots,b\}.$ 
For real number $c \in \mathbb{R}$, let $[c]^+ := \max\{c,0\}$.
For probability mass functions \(P\) and \(Q\) on a finite alphabet $\mathcal{X}$ such that $P \ll Q$, we define the \emph{KL divergence}, \emph{variational distance}, and \emph{chi-squared divergence}, respectively, by
\begin{align*}
\mathbb D(P\|Q)
\triangleq
\sum_xP(x)\log\frac{P(x)}{Q(x)},
\end{align*}
\begin{align*}
\mathbb V(P,Q)
\triangleq
\frac12\sum_x|P(x)-Q(x)|,
\end{align*}
\begin{align*}
\chi_2(P\|Q)
\triangleq
\sum_x\frac{(P(x)-Q(x))^2}{Q(x)}.
\end{align*}

\subsection{System Model}
Let $\mathcal{X}=\{0,1\}$ be the alphabet for channel input  and $(\mathcal{Y} , \mathcal{Z})$ be two other finite alphabets. Let $W_{Y|X}\in \mathcal{P}(\mathcal{Y}|\mathcal{X})$ be a conditional distribution mapping from $\mathcal{X}$ to $\mathcal{Y}$, and $W_{Z|X}\in \mathcal{P}(\mathcal{Z}|\mathcal{X})$ be a conditional distribution mapping from $\mathcal{X}$ to $\mathcal{Z}$. 
We consider a transmitter communicating with a legitimate receiver over the binary-input DMC \((\mathcal X,W_{Y|X},\mathcal Y)\), while a warden observes the channel input through a second DMC \((\mathcal X,W_{Z|X},\mathcal Z)\).  We refer to the tuple
\begin{align*}
(\mathcal X,W_{Y|X},W_{Z|X},\mathcal Y,\mathcal Z)
\end{align*}
as a \emph{binary-input covert communication channel} (as illustrated in Fig.~\ref{fig:system-model}). The symbol \(0\) denotes the \emph{innocent symbol}: when no communication happens, the transmitter sends the all-$0$ sequence. The symbol \(1\) denotes the \emph{non-innocent symbol} used to convey information. The output distributions at the legitimate receiver and the warden induced by the two input symbols are denoted by
\begin{align*}
P_0(y)\triangleq W_{Y|X}(y|0),
\qquad
P_1(y)\triangleq W_{Y|X}(y|1),
\end{align*}
and
\begin{align*}
Q_0(z)\triangleq W_{Z|X}(z|0),
\qquad
Q_1(z)\triangleq W_{Z|X}(z|1).
\end{align*}
Throughout the paper, we assume $
Q_1\ll Q_0, 
Q_1\ne Q_0,$ and
$P_1\ll P_0.$
The condition \(Q_1\ll Q_0\) excludes symbols that reveal communication with positive probability after a single use, whereas \(Q_1\ne Q_0\) excludes the degenerate case in which the warden obtains no statistical evidence of communication. The assumption \(P_1\ll P_0\) rules out the regime in which the legitimate parties can achieve a throughput that breaks the square-root law~\cite{Bloch2016,Wang2016}. For simplicity, we write
\begin{align*}
D_P\triangleq \mathbb D(P_1\|P_0),\qquad D_Q\triangleq \mathbb D(Q_1\|Q_0),\qquad
V_P\triangleq\operatorname{Var}_{P_1}\!\left[\log\frac{P_1(Y)}{P_0(Y)}\right].
\end{align*}
We impose the usual non-degeneracy conditions \(D_P>0\) and \(V_P>0\).

\begin{definition}
Let $M, K, n$ be three positive integers. An \((M,K,n)\) code contains an encoder and a decoder
\begin{align*}
f:\llbracket1,K\rrbracket\times\llbracket1,M\rrbracket
\longrightarrow\{0,1\}^n,
\qquad
\varphi:\mathcal Y^n\times\llbracket1,K\rrbracket
\longrightarrow\llbracket1,M\rrbracket.
\end{align*}
The message \(W \in \llbracket1,M\rrbracket\) and secret key \(S \in \llbracket1,K\rrbracket\)   are independent and uniform on their respective sets. The induced distribution of the warden's output, averaged over both message and key, satisfies that for each $\mathbf{z}\in\mathcal{Z}^n$
\begin{align*}
\widehat P_{Z^n}(\mathbf z)
\triangleq
\frac1{MK}\sum_{s=1}^K\sum_{w=1}^M
W_{Z|X}^{\otimes n}(\mathbf z|f(s,w)).
\end{align*}
The maximum probability of error is
\begin{align*}
P_{\mathrm{err}}^*
\triangleq
\max_{\substack{s\in\llbracket1,K\rrbracket, 
w\in\llbracket1,M\rrbracket}}
\Pr\{\varphi(Y^n,s)\ne w\mid S=s,W=w\}.
\end{align*}
\end{definition}

\begin{definition}
The code is \(\epsilon\)-reliable if \(P_{\mathrm{err}}^*\le\epsilon\).
An \(\epsilon\)-reliable code is an \((M,K,n,\epsilon,\delta)_{\mathbb V}\) code if
\begin{align*}
\mathbb V(\widehat P_{Z^n},Q_0^{\otimes n})\le\delta,
\end{align*}
and is an \((M,K,n,\epsilon,\delta)_{\mathbb{D}}\) code if
\begin{align*}
\mathbb{D} (\widehat P_{Z^n}\|Q_0^{\otimes n})\le\delta.
\end{align*}
Let \(M_{\mathbb{D}}^*(n,\epsilon,\delta)\) and \(M_{\mathbb V}^*(n,\epsilon,\delta)\) be the corresponding largest sizes of the message sets.
\end{definition}

\subsection{Pulse-Position Modulation}
 Let $m$ and $\ell$ be positive integers. A \emph{length-\(m\), weight-one PPM block} selects one coordinate uniformly over the $m$ coordinates, transmits \(1\) in that coordinate, and transmits \(0\) elsewhere. The resulting distribution of the warden's output is
\begin{align*}
P_m(\mathbf z)
\triangleq
\frac1m\sum_{j=1}^m
Q_1(z_j)\prod_{k\ne j}Q_0(z_k),
\end{align*}
Consider \(\ell\) independent PPM blocks, and let the total blocklength \(n=m\ell+r\), where \(0\le r<\ell\). Here, $\ell$ denotes the number of pulses (i.e., the number of ones in a codeword), and $r$ denotes the number of trailing zeros appended after the $\ell$ blocks. We define the \emph{PPM-induced output distribution} as 
\begin{align}
P_{Z,\mathrm{PPM}}^{n,\ell}
\triangleq
P_m^{\otimes\ell}\otimes Q_0^{\otimes r}.
\label{eq:ppm-output}
\end{align}
Let $c_0$ and $c_1$ be positive constants. The analysis in this work only considers the parameter regime
\begin{align}
0<c_0\le\frac\ell m\le c_1<\infty,
\label{eq:block-regime}
\end{align}
since both $\ell$ and $m$ scale as $\Theta(\sqrt{n})$ in the covert communication regime, as established in prior works~\cite{BlochGuha2017, Tahmasbi2019}.

\subsection{Main Results And Discussions}
For \(x\in\mathbb R\), let
 $Q(x)\triangleq\frac{1}{\sqrt{2\pi}}\int_x^\infty e^{-u^2/2}\,du$
denote the Gaussian \(Q\)-function. The following theorem provides the approximation of variational distance between the PPM-induced output distribution $P_{Z,\mathrm{PPM}}^{n,\ell}$ and the innocent distributions $Q_0^{\otimes n}$. 

\par\noindent{\bfseries Theorem 1 (Variational distance for the distribution induced by PPM).}
Suppose $n = m\ell + r$ such that $0<c_0\le\frac\ell m\le c_1<\infty$ and $0 \le r < \ell$. Then,  we have
\begin{align*}
\mathbb V(P_{Z,\mathrm{PPM}}^{n,\ell},Q_0^{\otimes n})
=
1-2Q\left(
\frac\ell2\sqrt{\frac{\chi_2(Q_1\|Q_0)}{n}}
\right)
+O(n^{-1/2}).
\end{align*}

\noindent{\textbf{Remark 1.}}
\emph{The approximation in Theorem~1 is sharper than the corresponding approximation in \cite[Lemma 8,~Eq.~(136)]{Tahmasbi2019}, which  has the same Gaussian leading term but contains the additional term \(2/\sqrt{\ell}\), together with an \(O(n^{-1/2})\) error term. Since \(\ell=\Theta(\sqrt n)\), the additional term is of order \(n^{-1/4}\). Theorem~1 removes this term by preserving the cancellation between the two likelihood-ratio contributions before applying the Gaussian approximation, and consequently yields a uniform error of order \(O(n^{-1/2})\). This allows us to obtain a tighter achievability bound that matches the converse bound of~\cite{Tahmasbi2019}, as shown below.}

After introducing Theorem~1, we are now ready to state the main result of this paper using the shorthand notation
$\Gamma
\triangleq
Q^{-1}\left(\frac{1-\delta}{2}\right)$.

\par\noindent{\bfseries Theorem 2 (Exact second-order asymptotics under variational distance).} Fix \(0<\epsilon<1\).  For every covertness parameter \(0<\delta<1\),
\begin{align}
   \log M_{\mathbb V}^*(n,\epsilon,\delta)
   =
   \frac{2\Gamma D_P}{\sqrt{\chi_2(Q_1\|Q_0)}}n^{1/2}
   -
   \sqrt{\frac{2\Gamma V_P}{\sqrt{\chi_2(Q_1\|Q_0)}}}
   Q^{-1}(\epsilon)n^{1/4}
   +O(\log n).
   \label{eq:second-order}
\end{align}

\noindent{\textbf{Remark 2.}}
\emph{Note that the right-hand side of~\eqref{eq:second-order} coincides with the converse bounds in \cite[Eqs. (29)]{Tahmasbi2019}.
Therefore, Theorem 2 completes the exact second-order characterization for the variational distance metric. Although the statement in Theorem 2 does not impose any constraint on the key size, the PPM random-coding construction in Section~\ref{sec:coding} can be implemented with
\begin{align*}
\log K
=
\frac{2\Gamma}{\sqrt{\chi_2(Q_1\|Q_0)}}
[D_Q-D_P]^+n^{1/2}
+o(n^{1/2}).
\end{align*}
In particular, no secret key is needed  whenever
\(D_Q < D_P\), i.e., whenever the warden's channel is less informative
than the legitimate parties' channel.
}

\noindent{\textbf{Remark 3.}}
\emph{The first- and second-order asymptotics under the KL-divergence metric were
characterized in \cite{Tahmasbi2019} as
\begin{align*}
\log M_{\mathbb{D}}^*(n,\epsilon,\delta)
=
\sqrt{\frac{2\delta}{\chi_2(Q_1\|Q_0)}}D_Pn^{1/2}
-
\sqrt{\sqrt{\frac{2\delta}{\chi_2(Q_1\|Q_0)}}V_P}\,
Q^{-1}(\epsilon)n^{1/4}
+O(\log n).
\end{align*}
This expression and~\eqref{eq:second-order} are both of the form
\(\ell D_P-\sqrt{\ell V_P}\,Q^{-1}(\epsilon)+O(\log n)\), where \(\ell\) is the
largest number of pulses imposed by the covertness constraint, that is,
\(\ell=\sqrt{2\delta n/\chi_2(Q_1\|Q_0)}\) under KL divergence and
\(\ell=2\Gamma\sqrt n/\sqrt{\chi_2(Q_1\|Q_0)}\) under variational distance. The
choice of covertness metric therefore affects only how the covertness parameter \(\delta\) is
converted into the largest number of pulses. Note that this differs qualitatively in the regime  \(\delta\to0\), where one has
\(\Gamma=\sqrt{\pi/2}\,\delta+O(\delta^3)\), so the number of pulses grows linearly in
\(\delta\) under variational distance, but grows as \(\sqrt{\delta}\) under KL divergence.
This makes intuitive sense because variational distance behaves like the square
root of KL divergence in this regime. Indeed, imposing \(\mathbb D\le2\delta^2\)
guarantees \(\mathbb V\le\delta\) by Pinsker's inequality and allows
\(2\delta\sqrt n/\sqrt{\chi_2(Q_1\|Q_0)}\) pulses, whereas working directly with
the variational distance allows \(\sqrt{2\pi}\,\delta\sqrt
n/\sqrt{\chi_2(Q_1\|Q_0)}\) pulses. The resulting gain in throughput, by a factor
of \(\sqrt{2\pi}/2\approx1.25\), is precisely the slack in Pinsker's inequality.}


\section{Sharp Approximation For Variational Distance (Proof of Theorem 1)}\label{sec:variational}

Recall from \eqref{eq:ppm-output} that the PPM-induced output distribution $P_{Z,\mathrm{PPM}}^{n,\ell}
=
P_m^{\otimes\ell}\otimes Q_0^{\otimes r}$,
whereas the innocent output distribution can be written as $
Q_0^{\otimes n}
=
\left(Q_0^{\otimes m}\right)^{\otimes\ell}
\otimes Q_0^{\otimes r}.$
Since tensoring two distributions with the same probability distribution
does not change their variational distance, we have
\begin{align*}
\mathbb V\left(
P_{Z,\mathrm{PPM}}^{n,\ell},
Q_0^{\otimes n}
\right)
=
\mathbb V\left(
P_m^{\otimes\ell},
\left(Q_0^{\otimes m}\right)^{\otimes\ell}
\right).
\end{align*}

\subsection{Midpoint Representation of Variational Distance}
We first convert the  variational distance $\mathbb V\left(
P_m^{\otimes\ell},
(Q_0^{\otimes m}\right)^{\otimes\ell}
)$ into two analytic quantities that can be estimated independently. For a length-$m$, weight one PPM block, we  define the \emph{Bhattacharyya coefficient} $\rho_m$ and the \emph{geometric midpoint distribution} $H_m$, respectively, by
\begin{align}
\rho_m
\triangleq
\sum_{\mathbf z}\sqrt{P_m(\mathbf z)Q_0^{\otimes m}(\mathbf z)},
\quad \text{and} \quad
H_m(\mathbf z)
\triangleq
\frac{\sqrt{P_m(\mathbf z)Q_0^{\otimes m}(\mathbf z)}}{\rho_m}. \label{rho}
\end{align}
Let \(\mathbf Z_1,\ldots,\mathbf Z_\ell\) be i.i.d. according to \(H_m\), and define the sum of log-likelihood ratio as 
\begin{align}
L_{m,\ell}
\triangleq
\sum_{i=1}^\ell
\log\frac{P_m(\mathbf Z_i)}{Q_0^{\otimes m}(\mathbf Z_i)}.  \label{eq:L}
\end{align}

\par\noindent{\bfseries Lemma 1.} By using the elementary  identity  
\begin{align*}
\min\{a,b\}
=
\sqrt{ab}
\exp\left(-\frac12\left|\log\frac ab\right|\right), \quad \forall a,b>0,
\end{align*}
the following expression holds:
\begin{align}
1-\mathbb V(P_m^{\otimes\ell},(Q_0^{\otimes m})^{\otimes\ell})
=
\rho_m^\ell \times 
\mathbb E_{H_m^{\otimes\ell}} 
\left[e^{-|L_{m,\ell}|/2}\right].
\label{eq:midpoint-overlap}
\end{align}
The proof of Lemma~1 is provided in Appendix~\ref{appendix:Lemma1}. Therefore, to approximate the variational distance,  it remains to evaluate the right-hand side of \eqref{eq:midpoint-overlap}. Let 
\begin{align}
h(x)\triangleq e^{-|x|/2}
\quad \text{and } \quad
N_{m,\ell}\sim
N\left(0,\frac{\ell\chi_2(Q_1\|Q_0)}{m}\right). \label{eq:N}
\end{align}
In what follows, we will prove that the following  two equations hold:
\begin{align*}
&\text{Equation (A):} \qquad
\rho_m^\ell
=
\exp\left[
-\frac{\ell\chi_2(Q_1\|Q_0)}{8m}
\right]
\left(1+O(m^{-1})\right),\\
&\text{Equation (B):}\qquad
\mathbb E_{H_m^{\otimes\ell}} \left(h(L_{m,\ell})\right)
=
\mathbb Eh(N_{m,\ell})
+O(m^{-1}).
\end{align*}
Equation (A) provides the required approximation of the power of Bhattacharyya coefficient $\rho_m^\ell$, whereas Equation (B) approximates the expectation of a function of $L_{m,\ell}$  by its Gaussian counterpart. In the rest of this section, Subsection~\ref{subsec:one-block} proves that  Equation (A) holds.  Subsection~\ref{subsec:fourier-reduction} converts the main
comparison in Equation (B) into the frequency domain and introduces an auxiliary Gaussian smoothing variable to prevent possible lattice peaks at high frequency.
Subsection~\ref{subsec:frequency-ranges} divides the frequency domain into four ranges and controls the error term
from each range. Subsection~\ref{subsec:desmoothing} removes the effect of the auxiliary Gaussian smoothing variable, thereby completing the proof of Equation (B).
Finally, Subsection~\ref{subsec:completion-tv} combines Equations (A)
and (B), evaluates the Gaussian expectation, and translates the
result from the block parameter $m$ to the original blocklength~\(n\).

\subsection{Approximation of the Power of Bhattacharyya Coefficient $\rho_m^{\ell}$}\label{subsec:one-block}
For brevity, we define
\begin{align}
B_m
\triangleq
\frac1m\sum_{j=1}^m
\left(
\frac{Q_1(Z_j)}{Q_0(Z_j)}-1
\right),
\label{eq:Bm-definition}
\end{align}
and Tahmasbi and Bloch~\cite{Tahmasbi2019} already calculated that when \((Z_1, Z_2, \ldots, Z_m) \sim Q_0^{\otimes m}\),
\begin{align*}
\mathbb E[B_m]=0,
\qquad
\mathbb E[B_m^2]=\frac{\chi_2(Q_1\|Q_0)}{m},
\qquad
\mathbb E[B_m^3]=O(m^{-2}),
\qquad
\mathbb E[B_m^4]=O(m^{-2}).
\end{align*}
Note that \(Q_1(Z)/Q_0(Z)-1\) is bounded on the finite support of \(Q_0\), hence we define
\begin{align*}
a_{\min}\triangleq
\min_{z:Q_0(z)>0}\left\{ \frac{Q_1(z)}{Q_0(z)}-1\right\}, \quad \text{and} \quad a_{\max}\triangleq
\max_{z:Q_0(z)>0} \left\{ \frac{Q_1(z)}{Q_0(z)}-1 \right\}.
\end{align*}
Hoeffding's inequality yields that
\begin{align}
Q_0^{\otimes m}\!\left(|B_m|>\frac12\right)
\le
2\exp\left(
-\frac{m}{2(a_{\max}-a_{\min})^2}
\right).
\label{eq:Bm-tail}
\end{align}

Recalling that $P_m(\mathbf z) = \frac1m\sum_{j=1}^m
Q_1(z_j)\prod_{k\ne j}Q_0(z_k)$, we have  \(P_m/Q_0^{\otimes m}=1+B_m\), and hence the Bhattacharyya coefficient can be written as $\rho_m = \mathbb E_{Q_0^{\otimes m}} [\sqrt{1+B_m}]$.
To evaluate $\rho_m$ through this expectation, we define the event
$\mathcal E_m
\triangleq
\left\{|B_m|\leq\frac12\right\}$, and then  decompose
\begin{align*}
\rho_m
=
\mathbb E_{Q_0^{\otimes m}}
\left[\sqrt{1+B_m}\mathbf 1_{\mathcal E_m}\right]
+
\mathbb E_{Q_0^{\otimes m}}
\left[\sqrt{1+B_m}\mathbf 1_{\mathcal E_m^{\mathrm c}}\right].
\end{align*}
We first consider the expectation over \(\mathcal E_m\). Let $
f(x)\triangleq\sqrt{1+x}$ for $x>-1$.
Taylor's theorem with the Lagrange remainder gives, for every
\(x\in[-1/2,1/2]\), there exists a real number $\xi_x \in [0,x]$ such that 
\begin{align*}
f(x) = \sqrt{1+x}
=
1+\frac{x}{2}-\frac{x^2}{8}+\frac{x^3}{16}
+R_4(x), \quad \text{where } R_4(x)
= \frac{f^{(4)}(\xi_x)}{24}x^4.
\end{align*}
For \(|x|\leq1/2\), we have \(\xi_x\geq-1/2\), and hence
$\left|f^{(4)}(\xi_x)\right|
\leq
\frac{15}{16}\left(\frac12\right)^{-7/2}.$
Therefore, there exists a constant \(c_2>0\) such that
$|R_4(x)|\leq c_2|x|^4$ for 
 $|x|\leq\frac12.$
Applying this expansion with \(x=B_m\) on event \(\mathcal E_m\) yields
\begin{align*}
\mathbb E_{Q_0^{\otimes m}}
\left[\sqrt{1+B_m}\mathbf 1_{\mathcal E_m}\right]
&=
Q_0^{\otimes m}(\mathcal E_m)
+\frac12
\mathbb E_{Q_0^{\otimes m}}
\left[B_m\mathbf 1_{\mathcal E_m}\right]
-\frac18
\mathbb E_{Q_0^{\otimes m}}
\left[B_m^2\mathbf 1_{\mathcal E_m}\right]\\
&\qquad\qquad
+\frac1{16}
\mathbb E_{Q_0^{\otimes m}}
\left[B_m^3\mathbf 1_{\mathcal E_m}\right]
+
O\left(
\mathbb E_{Q_0^{\otimes m}}
\left[B_m^4\mathbf 1_{\mathcal E_m}\right]
\right).
\end{align*}
It remains to remove the effect of the indicator \(\mathbf 1_{\mathcal E_m}\) in the above expression. Recalling the definitions of \(a_{\min}\) and \(a_{\max}\), we have
\begin{align*}
|B_m|
\leq
\max\{|a_{\min}|,|a_{\max}|\}
\qquad
Q_0^{\otimes m}\text{-almost surely}.
\end{align*}
Therefore, for \(k\in\{1,2,3\}\), the difference between $\mathbb E_{Q_0^{\otimes m}} \left[B_m^k\mathbf 1_{\mathcal E_m}\right]$ and  $\mathbb E_{Q_0^{\otimes m}}[B_m^k]$ is exponentially small, since  
\begin{align*}
\left|
\mathbb E_{Q_0^{\otimes m}}
\left[B_m^k\mathbf 1_{\mathcal E_m}\right]
-
\mathbb E_{Q_0^{\otimes m}}[B_m^k]
\right|
=
\left|
\mathbb E_{Q_0^{\otimes m}}
\left[B_m^k\mathbf 1_{\mathcal E_m^{\mathrm c}}\right]
\right| 
&\leq 
\max\{|a_{\min}|,|a_{\max}|\}^k
\times Q_0^{\otimes m}(\mathcal E_m^{\mathrm c}) \\
&=
O(e^{-(1/(2(a_{\max}-a_{\min})^2))m}),
\end{align*}
where the last step follows from~\eqref{eq:Bm-tail}.
Moreover, we also have $
\mathbb E_{Q_0^{\otimes m}}
\left[B_m^4\mathbf 1_{\mathcal E_m}\right]
\leq
\mathbb E_{Q_0^{\otimes m}}[B_m^4].$

We then consider the expectation over \(\mathcal E_m^{\mathrm c}\). Since \(1+B_m=P_m/Q_0^{\otimes m}\geq0\) and \(B_m\leq a_{\max}\), we have
$0\leq\sqrt{1+B_m}\leq\sqrt{1+a_{\max}}.$
Therefore,
\begin{align*}
\mathbb E_{Q_0^{\otimes m}}
\left[
\sqrt{1+B_m}\mathbf 1_{\mathcal E_m^{\mathrm c}}
\right]
&\leq
\sqrt{1+a_{\max}}\,
Q_0^{\otimes m}(\mathcal E_m^{\mathrm c}) =
O(e^{-(1/(2(a_{\max}-a_{\min})^2))m}),
\end{align*}
where the last step again follows from~\eqref{eq:Bm-tail}.
Combining the  analyses above yields that 
\begin{align*}
\rho_m
&=
1+\frac12\mathbb E_{Q_0^{\otimes m}}[B_m]
-\frac18\mathbb E_{Q_0^{\otimes m}}[B_m^2]
+\frac1{16}\mathbb E_{Q_0^{\otimes m}}[B_m^3]
+O\left(
\mathbb E_{Q_0^{\otimes m}}[B_m^4]
\right)
+O(e^{-(1/(2(a_{\max}-a_{\min})^2))m}) \\
&= 1-\frac{\chi_2(Q_1\|Q_0)}{8m}+O(m^{-2}),
\end{align*}
and hence
\begin{align*}
\log\rho_m
=
\log\left(1-\frac{\chi_2(Q_1\|Q_0)}{8m}+O(m^{-2})\right)
=
-\frac{\chi_2(Q_1\|Q_0)}{8m}+O(m^{-2}).
\end{align*}
Since \(1\leq\ell\leq c_1m\) from the condition of Theorem~1, multiplying $\ell$ on both sides yields
\begin{align*}
\ell\log\rho_m &=
-\frac{\ell\chi_2(Q_1\|Q_0)}{8m}
+O(m^{-1}).
\end{align*}
 Exponentiating both sides yields
\begin{align*}
\rho_m^\ell
&=
\exp\left(
-\frac{\ell\chi_2(Q_1\|Q_0)}{8m}
+O(m^{-1})
\right)\\
&=
\exp\left(
-\frac{\ell\chi_2(Q_1\|Q_0)}{8m}
\right)
\exp\left(O(m^{-1})\right)\\
&=
\exp\left(
-\frac{\ell\chi_2(Q_1\|Q_0)}{8m}
\right)
\left(1+O(m^{-1})\right).
\end{align*}
This completes the proof of Equation (A).


\subsection{Approximation of Equation (B) Via Fourier Representation In Characteristic Functions}\label{subsec:fourier-reduction}
 
 Our next objective is to approximate $\mathbb E [h(L_{m,\ell})]$. In particular, Subsections C--E below are organized to  prove that 
\begin{align}
\mathbb E [h(L_{m,\ell})] 
=
\mathbb E [h(N_{m,\ell})] 
+O(m^{-1}),
\label{eq:expectation-approximation}
\end{align}
where $L_{m,\ell}$ is defined in~\eqref{eq:L} and $N_{m,\ell}  \sim
N\left(0,\frac{\ell\chi_2(Q_1\|Q_0)}{m}\right)$ is defined in~\eqref{eq:N}.
To this end, we study the distributions of \(L_{m,\ell}\) and \(N_{m,\ell}\) through their
\emph{characteristic function}. Let \(i\) denote the imaginary unit. The characteristic function of  \(L_{m,\ell}\) under \(H_m^{\otimes\ell}\) is defined as
\begin{align*}
&\varphi_{L_{m,\ell}}(t)
\triangleq
\mathbb E_{H_m^{\otimes\ell}}
\left[e^{itL_{m,\ell}}\right],
\qquad t\in\mathbb R,
\end{align*}
whereas the characteristic function of  \(N_{m,\ell}\) is defined as
\begin{align*}
&\varphi_{N_{m,\ell}}(t)
\triangleq
\mathbb E \left[e^{itN_{m,\ell}}\right] = \exp\left(
-\frac{\ell\chi_2(Q_1\|Q_0)}{2m}t^2
\right),
\qquad t\in\mathbb R.
\end{align*}
This representation allows us to compare the distribution of
\(L_{m,\ell}\) with an appropriate Gaussian distribution $N_{m,\ell}$ by comparing
their characteristic functions over different frequency ranges.

For \(h(x)=e^{-|x|/2}\), its Fourier transform is
\begin{align*}
\widehat h(t)
=
\int_{\mathbb R}
e^{-|x|/2}e^{-itx}\,dx =
\frac{1}{t^2+1/4},
\qquad t\in\mathbb R.
\end{align*}
Since \(h\) is continuous and both \(h\) and \(\widehat h\) are
absolutely integrable, the Fourier inversion theorem yields
\begin{align*}
h(x)
= \frac{1}{2\pi}
\int_{\mathbb R}\widehat h(t)e^{itx}\,dt = 
\frac{1}{2\pi}
\int_{\mathbb R}
\frac{e^{itx}}{t^2+1/4}\,dt.
\end{align*}
Consequently, for every real-valued random variable \(X\) with characteristic function $\varphi_X(t)$,
Fubini's theorem yields
\begin{align}
\mathbb E[h(X)]
&=
\frac{1}{2\pi}
\int_{\mathbb R}
\frac{\mathbb E[e^{itX}]}
{t^2+1/4}\,dt = \frac{1}{2\pi}
\int_{\mathbb R}
\frac{\varphi_X(t)}
{t^2+1/4}\,dt, \label{eq:Fubini}
\end{align}
since $\mathbb E\int_{\mathbb R}
\frac{|e^{itX}|}{t^2+1/4}\,dt
=
\int_{\mathbb R}\frac{dt}{t^2+1/4}
<\infty.$

Directly applying~\eqref{eq:Fubini} to compare \(L_{m,\ell}\) and \(N_{m,\ell}\) through their characteristic functions, i.e., 
\begin{align}
\left| \mathbb E [h(L_{m,\ell})] - \mathbb E [h(N_{m,\ell})] \right| =  \frac{1}{2\pi} \left|
\int_{\mathbb R} 
\frac{\varphi_{L_{m,\ell}}(t) - \varphi_{N_{m,\ell}}(t)}
{t^2+1/4}\,dt \right| \label{eq:compare}
\end{align} 
would require controlling the integrand at arbitrarily high frequencies.
The Gaussian variable causes no difficulty, since
\(\varphi_{N_{m,\ell}}(t)\)
decays super-exponentially fast in \(|t|\). However, as  \(L_{m,\ell}\) is a sum of \(\ell\)
i.i.d. random variables taking finitely many values, its characteristic
function \(\varphi_{L_{m,\ell}}(t)\) may not be small and may return close to one
at  high frequencies, whereas the weight \((t^2+1/4)^{-1}\)
decays only quadratically. 

To circumvent this issue, we introduce a \emph{Gaussian smoothing variable} \(G_m\sim N(0,m^{-1})\), which is independent of \(L_{m,\ell}\) and \(N_{m,\ell} \). Its characteristic function is
\(\varphi_{G_{m}}(t) \triangleq \mathbb E [e^{itG_{m}}] = e^{-t^2/(2m)}\). By independence,
\begin{align*}
\varphi_{L_{m,\ell}+G_m}(t)
\triangleq  \mathbb E\left[e^{it(L_{m,\ell}+G_m)}\right]
&=
\varphi_{L_{m,\ell}}(t) \times e^{-t^2/(2m)},
\end{align*}
so that adding \(G_m\) multiplies \(\varphi_{L_{m,\ell}}\) by a factor of $e^{-t^2/(2m)}$, which is
negligible once $|t|$ exceeds the order of \(\sqrt{m\log m}\). This is precisely the range in which no useful
bound on \(\varphi_{L_{m,\ell}}\) is available.

After introducing $G_m$, we can then use the triangle inequality to decompose~\eqref{eq:compare} as follows: 
\begin{align}
\left|
\mathbb E [h(L_{m,\ell})]
-
\mathbb E [h(N_{m,\ell})]
\right|
&\le
\left|
\mathbb E [h(L_{m,\ell})]
-
\mathbb E [h(L_{m,\ell}+G_m)]
\right|
+
\left|
\mathbb E[h(L_{m,\ell}+G_m)]
-
\mathbb E[h(N_{m,\ell}+G_m)]
\right| \notag \\
&+
\left|
\mathbb E[h(N_{m,\ell}+G_m)]
-
\mathbb E[h(N_{m,\ell})]
\right|.\label{eq:desmoothing-decomposition}
\end{align}
In the following, we will analyze the \emph{smoothed Fourier error} $\left|
\mathbb E[h(L_{m,\ell}+G_m)]
-
\mathbb E[h(N_{m,\ell}+G_m)]
\right|$ in Subsection~D; while the other two terms, corresponding to \emph{desmoothing errors},   will be analyzed  in Subsection E.

\subsection{Upper Bounds On the Smoothed Fourier Error}\label{subsec:frequency-ranges}

Applying the Fourier representation in~\eqref{eq:Fubini}, we bound the smoothed Fourier error as 
\begin{align}
&\left|
\mathbb E [h(L_{m,\ell}+G_m)]
-
\mathbb E [h(N_{m,\ell}+G_m)]
\right| \notag \\
&=
\frac{1}{2\pi}
\left|
\int_{\mathbb R}
\frac{
\varphi_{L_{m,\ell}+G_m}(t)
-
\varphi_{N_{m,\ell}+G_m}(t)
}{
t^2+1/4
}\,dt
\right| \notag \\
&=
\frac{1}{2\pi}
\left|
\int_{\mathbb R}
\frac{e^{-t^2/(2m)}}{t^2+1/4}
\left[
\varphi_{L_{m,\ell}}(t)
-
e^{-\ell\chi_2(Q_1\|Q_0)t^2/(2m)}
\right]dt
\right| \label{eq:ind} \\
&\leq
\frac{1}{2\pi}
\int_{\mathbb R}
\frac{e^{-t^2/(2m)}}{t^2+1/4}
\left|
\varphi_{L_{m,\ell}}(t)
-
e^{-\ell\chi_2(Q_1\|Q_0)t^2/(2m)}
\right|dt, \label{eq:smoothed-fourier-bound}
\end{align}
where~\eqref{eq:ind} follows since $\varphi_{N_{m,\ell}+G_m}(t) = \varphi_{N_{m,\ell}}(t) \times \varphi_{G_m}(t) = e^{-t^2/(2m)} \times e^{-\ell\chi_2(Q_1\|Q_0)t^2/(2m)}$. 
For simplicity, we define the \emph{nonnegative Fourier-error density} as 
\begin{align*}
\mathcal E_{m,\ell}(t)
\triangleq
\frac{e^{-t^2/(2m)}}{t^2+1/4}
\left|
\varphi_{L_{m,\ell}}(t)
-e^{-\ell\chi_2(Q_1\|Q_0)t^2/(2m)}
\right|.
\end{align*}
Fix \(k\ge 1\) and \(d\ge d_0\), where \(d_0>0\) is the constant specified in
Proposition~2 below, and let \(u_0>0\) be a constant detailed in
Appendix~\ref{appendix:proposition2}. Below, we decompose the integral in~\eqref{eq:smoothed-fourier-bound} into four frequency bands:
\begin{align}
\int_{\mathbb R}\mathcal E_{m,\ell}(t)\,dt
={}&
\int_{|t|\le d\sqrt{\log m}}\mathcal E_{m,\ell}(t)\,dt \notag \\
&+
\int_{d\sqrt{\log m}<|t|\le u_0\sqrt m}
\mathcal E_{m,\ell}(t)\,dt \notag \\
&+
\int_{u_0\sqrt m<|t|\le\sqrt{km\log m}}
\mathcal E_{m,\ell}(t)\,dt \notag \\
&+
\int_{|t|>\sqrt{km\log m}}
\mathcal E_{m,\ell}(t)\,dt. \label{eq:decom}
\end{align}
The four frequency bands arise because different estimates of \(\varphi_{L_{m,\ell}}(t)\)
are available at different frequencies, and each estimate is valid only up to a
certain frequency.
\begin{itemize}
\item 
On \(|t|\le d\sqrt{\log m}\), a refined expansion of \(\varphi_{L_{m,\ell}}(t)\)
is  accurate enough to control the \emph{difference}
\(\varphi_{L_{m,\ell}}(t)-e^{-\ell\chi_2(Q_1\|Q_0)t^2/(2m)}\) directly. This
band is what determines the \(O(m^{-1})\) rate in Equation~(B); the other three
bands are only required not to exceed it, and are in fact smaller once \(d\)
and \(k\) are chosen large enough.

\item On \(d\sqrt{\log m}<|t|\le u_0\sqrt m\), this expansion is no longer accurate,
so we bound \(\varphi_{L_{m,\ell}}(t)\) and
\(e^{-\ell\chi_2(Q_1\|Q_0)t^2/(2m)}\) separately by the triangle inequality.
Both decay as \(e^{-\Theta(t^2)}\) on this band, and integrating this decay
over \(|t|>d\sqrt{\log m}\) leads to \(m^{-\Theta(d^2)}\).

\item On \(u_0\sqrt m<|t|\le\sqrt{km\log m}\), even this quadratic decay is
unavailable. What remains is that the characteristic function of a single
summand of \(L_{m,\ell}\) in~\eqref{eq:L} stays bounded away
from one, uniformly over this band; raised to the \(\ell\)-th power, it becomes
\(e^{-\Theta(m)}\). The Gaussian term is exponentially small here as well.

\item On \(|t|>\sqrt{km\log m}\), no estimate of \(\varphi_{L_{m,\ell}}\) is used at
all: the smoothing factor \(e^{-t^2/(2m)}\) is at most \(m^{-k/2}\) there,
which already makes the integral negligible. This is precisely the band in
which lattice effects could keep \(\varphi_{L_{m,\ell}}\) from being small, and
it is the reason for introducing \(G_m\).
\end{itemize}
Propositions~1--4 below make these four estimates precise.

\par\noindent{\bfseries Proposition 1.} For every fixed \(d>0\),
\begin{align*}
\int_{|t|\le d\sqrt{\log m}}
\mathcal E_{m,\ell}(t)\,dt
=O(m^{-1}).
\end{align*}

\par\noindent{\bfseries Proposition 2.} There exist \(u_0>0\) and \(d_0>0\) such that, for every fixed \(d\ge d_0\),
\begin{align*}
\int_{d\sqrt{\log m}<|t|\le u_0\sqrt m}
\mathcal E_{m,\ell}(t)\,dt
=O(m^{-1}).
\end{align*}

\par\noindent{\bfseries Proposition 3.}
For every fixed \(k>0\), there exists a constant \(c_{3,k}>0\) such that
\begin{align}
\int_{u_0\sqrt m<|t|\leq\sqrt{km\log m}}
\mathcal E_{m,\ell}(t)\,dt
=
O(e^{-c_{3,k}m}),
\label{eq:contraction-range}
\end{align}
where \(u_0>0\) is the constant chosen in Proposition 2.

\par\noindent{\bfseries Proposition 4.} For every fixed \(k\ge1\),
\begin{align}
\int_{|t|>\sqrt{km\log m}}
\mathcal E_{m,\ell}(t)\,dt
=O(m^{-1}).
\label{eq:smoothing-tail}
\end{align}

The proofs of Propositions 1-4 are provided in Appendices~\ref{appendix:proposition1}-\ref{appendix:proposition4}, respectively. Combining~\eqref{eq:smoothed-fourier-bound},~\eqref{eq:decom}, and Propositions 1-4 together, it can be shown that the smoothed Fourier error satisfies
\begin{align*}
|\mathbb E [h(L_{m,\ell}+G_m)] - \mathbb E [h(N_{m,\ell}+G_m)] | = O(m^{-1}).
\end{align*}


\subsection{Removal of Gaussian Smoothing}
\label{subsec:desmoothing}

To complete the decomposition in \eqref{eq:desmoothing-decomposition}, it remains to show that removing \(G_m\) from each of the two expectations does not incur significant error. 

\medskip
\par\noindent{\bfseries Proposition 5 (Desmoothing error).}
The two desmoothing error terms satisfy
\begin{align*}
\left|
\mathbb E[h(L_{m,\ell}+G_m)]-\mathbb E[h(L_{m,\ell})]
\right|
=O(m^{-1}),
\qquad
\left|
\mathbb E[h(N_{m,\ell}+G_m)]-\mathbb E[h(N_{m,\ell})]
\right|
=O(m^{-1}).
\end{align*}

The proof of Proposition 5 is presented in Appendix~\ref{app:desmoothing}. Below, we briefly
describe the high-level intuition for~\(L_{m,\ell}\). The argument for $N_{m,\ell}$ is similar. 

The key observation is the almost-sure perturbation bound
\begin{align}
\left|
h(L_{m,\ell}+G_m)-h(L_{m,\ell})
-G_mh'(L_{m,\ell})
\right|
\le
G_m^2+
|G_m|\mathbf 1_{\{|L_{m,\ell}|\le |G_m|\}}. \label{eq:pro5}
\end{align}
The quadratic term $G_m^2$ controls the usual second-order Taylor error when
the interval between \(L_{m,\ell}\) and \(L_{m,\ell}+G_m\) does not
contain the origin. If this interval crosses the origin, then
\(h(t)=e^{-|t|/2}\) is not differentiable along the interval, and the error
may instead be of order \(|G_m|\). Such a crossing can occur only if
\(|L_{m,\ell}|\le |G_m|\), which accounts for the second term in the
bound.

Because \(G_m\) is independent of \(L_{m,\ell}\) and has zero mean,
the linear term \(G_mh'(L_{m,\ell})\) has zero expectation. It
therefore remains to control the two terms on the right-hand side of~\eqref{eq:pro5}.
The first quadratic term contributes
\(\mathbb E[G_m^2]=m^{-1}\). For the second term, the
anti-concentration estimate established in Appendix~\ref{app:desmoothing} gives
\(\Pr\{|L_{m,\ell}|\le s\}\le c(s+m^{-1/2})\) uniformly over
\(s\ge0\). Conditioning on \(G_m\) and taking \(s=|G_m|\) yields that
the second term is bounded by
\(c\,\mathbb E[|G_m|(|G_m|+m^{-1/2})]\), which is \(O(m^{-1})\)
since \(\mathbb E[G_m^2]=m^{-1}\) and
\(\mathbb E|G_m|=O(m^{-1/2})\). Hence adding and subsequently
removing the smoothing variable changes the expectation by only
\(O(m^{-1})\).

Combining Subsections D and E together, we prove that $\mathbb E [h(L_{m,\ell})] 
=
\mathbb E [h(N_{m,\ell})] 
+O(m^{-1})$ in~\eqref{eq:expectation-approximation}.

\subsection{Completion of the Proof of Theorem 1}\label{subsec:completion-tv}

Substituting Equation (A) and \eqref{eq:expectation-approximation} into Lemma 1
gives
\begin{align}
1-\mathbb V\left(
P_m^{\otimes\ell},
(Q_0^{\otimes m})^{\otimes\ell}
\right)
&=
\rho_m^\ell \times 
\mathbb E_{H_m^{\otimes\ell}}
\left[h(L_{m,\ell})\right]
\notag \\
&\quad=
\exp\left(
-\frac{\ell\chi_2(Q_1\|Q_0)}{8m}
\right)
\left(1+O(m^{-1})\right)
\left(
\mathbb E [h(N_{m,\ell})]+O(m^{-1})
\right)
\notag \\
&\quad=
\exp\left(
-\frac{\ell\chi_2(Q_1\|Q_0)}{8m}
\right)
\mathbb E [h(N_{m,\ell})]
+
O(m^{-1}), \label{eq:haha} 
\end{align}
where the last step uses \(0\leq h(\cdot )\leq1\) and the fact that \(\ell/m\) is
bounded (as stated in~\eqref{eq:block-regime}).

It remains to evaluate $\mathbb E[ h(N_{m,\ell})]$ where $N_{m,\ell} \sim
N\left(0,\frac{\ell\chi_2(Q_1\|Q_0)}{m}\right)$. By symmetry and the
density of \(N_{m,\ell}\),
\begin{align}
\mathbb E [h(N_{m,\ell})]=
\mathbb E e^{-|N_{m,\ell}|/2}
&=
2\sqrt{
\frac{m}{
2\pi\ell\chi_2(Q_1\|Q_0)
}
}
\int_0^\infty
\exp\left(
-\frac{x}{2}
-\frac{mx^2}{
2\ell\chi_2(Q_1\|Q_0)
}
\right)dx \notag \\
&=
2\exp\left(
\frac{\ell\chi_2(Q_1\|Q_0)}{8m}
\right)
\sqrt{
\frac{m}{
2\pi\ell\chi_2(Q_1\|Q_0)
}
} \notag \\
&\qquad\times
\int_0^\infty
\exp\left(
-\frac{m}{2\ell\chi_2(Q_1\|Q_0)}
\left(
x+\frac{\ell\chi_2(Q_1\|Q_0)}{2m}
\right)^{2}
\right)dx \notag \\
&=2\exp\left(
\frac{\ell\chi_2(Q_1\|Q_0)}{8m}
\right)
Q\left(
\frac12
\sqrt{
\frac{\ell\chi_2(Q_1\|Q_0)}{m}
}
\right), \label{eq:gaussian-overlap}
\end{align}
where the last step follows from the change of variable
$u=\sqrt{\frac{m}{\ell\chi_2(Q_1\|Q_0)}}\left(x+\frac{\ell\chi_2(Q_1\|Q_0)}{2m}\right)$,
under which the lower limit $x=0$ is mapped to
$u=\frac12\sqrt{\frac{\ell\chi_2(Q_1\|Q_0)}{m}}$.
Substituting \eqref{eq:gaussian-overlap} into the expression in~\eqref{eq:haha}, the two exponential
factors cancel, and hence
\begin{align*}
1-\mathbb V\left(
P_m^{\otimes\ell},
(Q_0^{\otimes m})^{\otimes\ell}
\right)
=
2Q\left(
\frac12
\sqrt{
\frac{\ell\chi_2(Q_1\|Q_0)}{m}
}
\right)
+
O(m^{-1}).
\end{align*}
It remains to replace the argument of \(Q\) from $\frac12
\sqrt{\frac{\ell\chi_2(Q_1\|Q_0)}{m}}$ to $\frac{\ell}{2} \sqrt{\frac{\chi_2(Q_1\|Q_0)}{n}}$  that appeared  in
Theorem 1. Recall that  \(n=m\ell+r\). The two arguments coincide exactly when \(r=0\). When $r \ne 0$, the following computation shows that the non-zero remainder \(r\)
perturbs the argument by only \(O(m^{-1})\).  Note that 
\begin{align*}
\left[
\frac12
\sqrt{
\frac{\ell\chi_2(Q_1\|Q_0)}{m}
}
\right]^2
-
\left[
\frac{\ell}{2}
\sqrt{
\frac{\chi_2(Q_1\|Q_0)}{n}
}
\right]^2 =
\frac{\ell\chi_2(Q_1\|Q_0)}{4m}
\left(
1-\frac{m\ell}{n}
\right)=
\frac{\ell\chi_2(Q_1\|Q_0)}{4m}
\frac{r}{n}.
\end{align*}
Because \(r<\ell\), \(n\geq m\ell\), and \(\ell/m\leq c_1\), we have
\begin{align*}
0
\leq
\frac{\ell\chi_2(Q_1\|Q_0)}{4m}\frac{r}{n}
\leq
\frac{\ell\chi_2(Q_1\|Q_0)}{4m}\frac{\ell}{m\ell}\leq
\frac{c_1\chi_2(Q_1\|Q_0)}{4m}=
O(m^{-1}).
\end{align*}
Using the fact that 
$|x-y|
=
\frac{|x^2-y^2|}{x+y}$
for $x,y\geq0$, we have 
\begin{align*}
\left| 
\frac12
\sqrt{
\frac{\ell\chi_2(Q_1\|Q_0)}{m}
}
-
\frac{\ell}{2}
\sqrt{
\frac{\chi_2(Q_1\|Q_0)}{n}
} \right| = O(m^{-1}).
\end{align*}
Since
$|Q'(x)|
=
\frac{1}{\sqrt{2\pi}}e^{-x^2/2}
\leq
\frac{1}{\sqrt{2\pi}},$
the function \(Q\) is Lipschitz, and therefore
\begin{align*}
&
Q\left(
\frac12
\sqrt{
\frac{\ell\chi_2(Q_1\|Q_0)}{m}
}
\right)=
Q\left(
\frac{\ell}{2}
\sqrt{
\frac{\chi_2(Q_1\|Q_0)}{n}
}
\right)
+
O(m^{-1}).
\end{align*}
Finally, combining the preceding calculations and noting that $m = \Theta(\sqrt{n})$, we obtain 
\begin{align*}
\mathbb V\left(
P_{Z,\mathrm{PPM}}^{n,\ell},
Q_0^{\otimes n}
\right)
=
1-
2Q\left(
\frac{\ell}{2}
\sqrt{
\frac{\chi_2(Q_1\|Q_0)}{n}
}
\right)
+
O(n^{-1/2}).
\end{align*}
This completes the proof of Theorem~1.

\section{Derivation of the Exact Second-Order Coding Asymptotics}\label{sec:coding}
This section links the approximation of variational distance in Section III to the covert communication problem.

\subsection{Random Coding With PPM Input Distribution}
Recall that \(n=m\ell+r\), where \(0\le r<\ell\), and we partition the first \(m\ell\) coordinates into the blocks
\begin{align*}
\mathcal B_i\triangleq\llbracket (i-1)m+1,im\rrbracket,
\qquad i\in\llbracket1,\ell\rrbracket.
\end{align*}
The length-\(n\), weight-\(\ell\) \emph{PPM input distribution} is uniform on the set
\begin{align*}
\mathcal P_{n,\ell}
\triangleq
\left\{
\mathbf x\in\{0,1\}^n:
\sum_{j\in\mathcal B_i}x_j=1\ \text{for every }i,
\quad
\sum_{j=m\ell+1}^{n}x_j=0
\right\}.
\end{align*}
Thus every codeword has one non-innocent symbol in each of the \(\ell\) blocks and only innocent symbols in the remaining $r$ coordinates. There are \(m^\ell\) such sequences, each having probability \(m^{-\ell}\). Passing this input distribution through the warden's channel gives exactly the PPM-induced output distribution \(P_{Z,\mathrm{PPM}}^{n,\ell}\) in \eqref{eq:ppm-output}.

For message size \(M_n\) and key size \(K_n\), we  draw codewords
$
\bigl\{\mathbf X_{sw}:s\in\llbracket1,K_n\rrbracket,
w\in\llbracket1,M_n\rrbracket\bigr\}$
independently from the PPM input distribution. For a fixed key \(s\), the subcode \(\{\mathbf X_{sw}:w\in\llbracket1,M_n\rrbracket\}\) is used for reliable communication to the legitimate receiver. The decoding rule is the same as that in~\cite{Tahmasbi2019}.

The reliability, resolvability, and expurgation
arguments for this random codebook were established in
\cite[Theorem 5]{Tahmasbi2019} and will not be repeated here.
We only focus on the specialization of that result needed in the
present paper. This specialization also makes explicit that the
only new ingredient required below is a sufficiently accurate
estimate of the covertness of the PPM-induced output distribution.

\medskip
\par\noindent{\bfseries Proposition 6.}
Fix \(0<\epsilon<1\), and let
\(\ell=\Theta(\sqrt n)\). Suppose that the PPM-induced output
distribution satisfies
\begin{align*}
\mathbb V\bigl(P_{Z,\mathrm{PPM}}^{n,\ell},
Q_0^{\otimes n}\bigr)
\le \delta-n^{-1/2}.
\end{align*}
Then, for all sufficiently large \(n\), there exists a deterministic
PPM code satisfying the corresponding covertness constraint and
having maximum probability of error at most \(\epsilon\), with
\begin{align*}
\log M_n
=
\ell D_P-\sqrt{\ell V_P}\,Q^{-1}(\epsilon)+O(\log n), \text{ and } 
\log K_n
=
\ell[D_Q-D_P]^++o(\sqrt n).
\end{align*}

The proof of Proposition 6 is provided in Appendix~\ref{appendix:proposition6}, where we verify that the four  conditions required for ~\cite[Theorem 5]{Tahmasbi2019} to hold are satisfied under  the length-\(n\),
weight-\(\ell\) PPM input distribution $\mathcal P_{n,\ell}$.

\subsection{Pulse Selection under the Variational Distance Metric}

Fix \(0<\delta<1\), and recall that $\Gamma
=
Q^{-1}\left(\frac{1-\delta}{2}\right).$
By Theorem 1, there exists a constant \(c_4>0\)  such that
\begin{align*}
\mathbb V\bigl(
P_{Z,\mathrm{PPM}}^{n,\ell},Q_0^{\otimes n}
\bigr)
\le
1-2Q\left(
\frac{\ell}{2}
\sqrt{\frac{\chi_2(Q_1\|Q_0)}{n}}
\right)
+\frac{c_4}{\sqrt n}
\end{align*}
for all sufficiently large \(n\). Choose
\begin{align*}
\ell_n
=
\left\lfloor
\frac{2\Gamma}{\sqrt{\chi_2(Q_1\|Q_0)}}\sqrt n-c_5
\right\rfloor,
\end{align*}
where \(c_5>0\) is a constant satisfying $\frac{c_5\sqrt{\chi_2(Q_1\|Q_0)}}
{\sqrt{2\pi}\,e^{\Gamma^2/2}}
>
c_4+1
$. By writing 
$\ell_n
=
\frac{2\Gamma}{\sqrt{\chi_2(Q_1\|Q_0)}}\sqrt n
-c_5-\theta_n$ for 
$0\le\theta_n<1$, we have
\begin{align*}
\frac{\ell_n}{2}
\sqrt{\frac{\chi_2(Q_1\|Q_0)}{n}}
=
\Gamma
-
\frac{(c_5+\theta_n)\sqrt{\chi_2(Q_1\|Q_0)}}
{2\sqrt n}.
\end{align*}
Since \(1-2Q(\Gamma)=\delta\), Taylor expansion at \(\Gamma\)
yields
\begin{align*}
1-2Q\left(
\frac{\ell_n}{2}
\sqrt{\frac{\chi_2(Q_1\|Q_0)}{n}}
\right)
&=
\delta
-
\frac{(c_5+\theta_n)\sqrt{\chi_2(Q_1\|Q_0)}}
{\sqrt{2\pi n}\,e^{\Gamma^2/2}}
+O(n^{-1})                                                   \\
&\le
\delta
-
\frac{c_5\sqrt{\chi_2(Q_1\|Q_0)}}
{\sqrt{2\pi n}\,e^{\Gamma^2/2}}
+O(n^{-1}).
\end{align*}
Consequently,
\begin{align*}
\mathbb V\bigl(
P_{Z,\mathrm{PPM}}^{n,\ell_n},Q_0^{\otimes n}
\bigr)
\le
\delta
-
\frac{1}{\sqrt n}
\left(
\frac{c_5\sqrt{\chi_2(Q_1\|Q_0)}}
{\sqrt{2\pi}\,e^{\Gamma^2/2}}
-c_4+o(1)
\right).
\end{align*}
Since the constant $c_5$ satisfies $\frac{c_5\sqrt{\chi_2(Q_1\|Q_0)}}
{\sqrt{2\pi}\,e^{\Gamma^2/2}}
>
c_4+1,$ we have that  for all sufficiently large \(n\),
\begin{align*}
\mathbb V\bigl(
P_{Z,\mathrm{PPM}}^{n,\ell_n},Q_0^{\otimes n}
\bigr)
\le
\delta-\frac1{\sqrt n},
\end{align*}
and hence the condition of Proposition 6 is satisfied.

Proposition 6 therefore gives a sequence of
\((M_n,K_n,n,\epsilon,\delta)_{\mathbb V}\) codes satisfying
\begin{align*}
\log M_n
=
\ell_nD_P
-
\sqrt{\ell_nV_P}\,Q^{-1}(\epsilon)
+
O(\log n).
\end{align*}
Substitution $
\ell_n
=
\frac{2\Gamma}{\sqrt{\chi_2(Q_1\|Q_0)}}\sqrt n+O(1)$ therefore  yields
\begin{align*}
\log M_n
=
\frac{2\Gamma D_P}
{\sqrt{\chi_2(Q_1\|Q_0)}}n^{1/2}
-
\sqrt{
\frac{2\Gamma V_P}
{\sqrt{\chi_2(Q_1\|Q_0)}}
}\,
Q^{-1}(\epsilon)n^{1/4}
+
O(\log n)
.
\end{align*}
The \(n^{1/4}\) penalty appearing in \cite[Eq. (30)]{Tahmasbi2019} has therefore disappeared. Finally, by combining the converse result for variational distance in \cite{Tahmasbi2019}, we complete the proof of Theorem 2. 

\section{Conclusion}

This paper has established the exact second-order asymptotics for covert communication over the considered binary-input DMCs when covertness is measured by variational distance. The coding scheme is the PPM construction used in previous work, whereas the improvement mainly comes from a sharper analysis of the distribution induced at the warden.  The resulting \(O(n^{-1/2})\) approximation eliminates the \(O(n^{1/4})\) backoff that appeared in the previous achievability analysis. Consequently, the achievable second-order term coincides with the converse second-order term developed in \cite{Tahmasbi2019}.

A promising direction for future work is to investigate whether the techniques developed here can be adapted to the setting in which covertness is measured by the probability of missed detection at a fixed probability of false alarm. The corresponding criterion is governed by a Neyman--Pearson tradeoff and has a different structure from variational distance, so it remains unclear whether the refinement established in this paper carries over without additional losses. A successful extension could lead to matching second-order asymptotics for this metric.

\section*{Acknowledgement}

The proofs in this paper are provided jointly by the authors and GPT-5.6-Sol, with support from a math harness project at Shanghai AI Laboratory, developed by Dr. Shuyue Hu, Dr. Yang Chen, and Mr. Zhanhao Li. The  authors performed a complete verification of the part of proofs that are generated by AI, including all intermediate calculations and logical steps, and assume full responsibility for the correctness of the mathematical content.

\appendices

\section{Proof of Lemma 1}  \label{appendix:Lemma1}
For arbitrary probability mass functions \(R\) and \(S\),
\begin{align*}
1-\mathbb V(R,S)
&=
\frac12\sum_x\{R(x)+S(x)-|R(x)-S(x)|\}\\
&=
\sum_x\min\{R(x),S(x)\}.
\end{align*}
Substituting \(a=P_m^{\otimes\ell}\), \(b=(Q_0^{\otimes m})^{\otimes\ell}\), and writing \(\underline{\mathbf z}=(\mathbf z_1,\ldots,\mathbf z_\ell)\), at every $\underline{\mathbf z}$ for which both  distributions $P_m^{\otimes\ell}(\underline{\mathbf z})$ and $(Q_0^{\otimes m})^{\otimes\ell}(\underline{\mathbf z})$  are positive, we have
\begin{align*}
\min\left\{
P_m^{\otimes\ell}(\underline{\mathbf z}),
(Q_0^{\otimes m})^{\otimes\ell}(\underline{\mathbf z})
\right\} =
\sqrt{
P_m^{\otimes\ell}(\underline{\mathbf z})
(Q_0^{\otimes m})^{\otimes\ell}(\underline{\mathbf z})
}
\exp\left[
-\frac12
\left|
\log\frac{
P_m^{\otimes\ell}(\underline{\mathbf z})
}{
(Q_0^{\otimes m})^{\otimes\ell}(\underline{\mathbf z})
}
\right|
\right].
\end{align*}
By the definition of \(H_m\), we have \(\sqrt{P_m(\mathbf z_i)Q_0^{\otimes m}(\mathbf z_i)}=\rho_m H_m(\mathbf z_i)\) and thus
\begin{align*}
\sqrt{
P_m^{\otimes\ell}(\underline{\mathbf z})
(Q_0^{\otimes m})^{\otimes\ell}(\underline{\mathbf z})
}
=
\rho_m^\ell
\prod_{i=1}^{\ell}H_m(\mathbf z_i) = \rho_m^\ell H_m^{\otimes\ell}(\underline{\mathbf z}).
\end{align*}
Similarly, the logarithm of the product likelihood ratio can be expressed as
\begin{align*}
\log\frac{
P_m^{\otimes\ell}(\underline{\mathbf z})
}{
(Q_0^{\otimes m})^{\otimes\ell}(\underline{\mathbf z})
}
=
\sum_{i=1}^{\ell}
\log\frac{P_m(\mathbf z_i)}{Q_0^{\otimes m}(\mathbf z_i)} =   L_{m,\ell}(\underline{\mathbf z}). 
\end{align*}
If at $\underline{\mathbf z}$ one of the product masses is zero, its contribution to $\min\{ P_m^{\otimes\ell}, (Q_0^{\otimes m})^{\otimes\ell} \} $ and $ H_m^{\otimes\ell}$ are both zero. Consequently, 
\begin{align*}
1-\mathbb V(P_m^{\otimes\ell},(Q_0^{\otimes m})^{\otimes\ell}) &= \sum_{\underline{\mathbf z}} \min\left\{
P_m^{\otimes\ell}(\underline{\mathbf z}),
(Q_0^{\otimes m})^{\otimes\ell}(\underline{\mathbf z})
\right\} \\
&= \sum_{\underline{\mathbf z}: H_m^{\otimes\ell}(\underline{\mathbf z}) > 0 }  \rho_m^\ell H_m^{\otimes\ell}(\underline{\mathbf z})
\exp\left(-\frac12|L_{m,\ell}(\underline{\mathbf z})|\right) =  \rho_m^\ell \times 
\mathbb E_{H_m^{\otimes\ell}}
\left[e^{-|L_{m,\ell}|/2}\right].
\end{align*} 
This completes the proof of Lemma~1.

\section{Proof of Proposition 1} \label{appendix:proposition1}

\par\noindent{\bfseries Proposition 1.} For every fixed \(d>0\),
\begin{align*}
\int_{|t|\le d\sqrt{\log m}}
\mathcal E_{m,\ell}(t)\,dt
=O(m^{-1}).
\end{align*}

Before introducing the proof of Proposition 1, we first introduce an auxiliary variable $M_m(s)$ that will be useful in later analyses:
\begin{align*}
M_m(s)
\triangleq
\mathbb E_{Q_0^{\otimes m}}
\left[(1+B_m)^s\right],
\qquad 
s\in\mathbb C \ \text{such that}\ \operatorname{Re}(s)>0,
\end{align*}
where \(\mathbb C\) represents the set of complex numbers, whereas 
\(\operatorname{Re}(s)\) denotes the real part of \(s\). Note that $\rho_m$ defined in~\eqref{rho} is a special case of $M_m(s)$ when $s = 1/2$.

\par\noindent{\bfseries Lemma 2.}
For every fixed \(d>0\), there exists a constant \(c_{B,1,d}>0\),
depending only on \(d\), \(Q_0\), and \(Q_1\), such that, for all
sufficiently large \(m\), uniformly over \(s\in\mathbb C\) satisfying
\begin{align*}
\frac14\leq\operatorname{Re}(s)\leq\frac34,
\qquad
|\operatorname{Im}(s)|\leq d\sqrt{\log m},
\end{align*}
one has
\begin{align*}
\log M_m(s)
=
\frac{\chi_2(Q_1\|Q_0)}{2m}s(s-1)+r_m(s),
\quad \text{where }
|r_m(s)|
\leq 
\frac{c_{B,1,d}(1+|s|)^4}{m^2}.
\end{align*}
Below, we first provide a proof for Lemma~2, and then present the proof of Proposition~1 based on Lemma~2.

\subsection{Proof of Lemma 2}\label{app:lemma2}

For convenience, we define
$A(Z)\triangleq\frac{Q_1(Z)}{Q_0(Z)}-1$. Then the random variable
\(B_m\) defined in \eqref{eq:Bm-definition} can be written as
$B_m=\frac1m\sum_{j=1}^m A(Z_j).$
Since \(Q_1\) and \(Q_0\) are probability mass functions, we have
\begin{align*}
\mathbb E_{Q_0}[A(Z)]
=0, \quad \text{and} \quad
\mathbb E_{Q_0}[A(Z)^2]
=
\chi_2(Q_1\|Q_0).
\end{align*}
Moreover, \(A(Z)\) is bounded because the alphabet is finite and
\(Q_1\ll Q_0\).

\medskip
\subsubsection{Calculating the Moments of \(B_m\)}
First, we note that~\cite[Eqns.~(359)--(363)]{Tahmasbi2019} has already shown that
\begin{align*}
\mathbb E B_m=0, \quad \mathbb E B_m^2 = \frac{\chi_2(Q_1\|Q_0)}{m}, \quad \mathbb E B_m^3 = \frac{\mathbb E_{Q_0}[A(Z)^3]}{m^2}.
\end{align*}
For the fourth moment, the only nonzero index patterns are four equal
indices and two pairs. It follows that
\begin{align*}
\mathbb E
\left[
\left(
\sum_{j=1}^m A(Z_j)
\right)^4
\right]
={}&
m\mathbb E_{Q_0}[A(Z)^4] +
6\sum_{1\le i<j\le m}
\mathbb E_{Q_0}[A(Z_i)^2]
\mathbb E_{Q_0}[A(Z_j)^2]\\
={}&
m\mathbb E_{Q_0}[A(Z)^4]
+
3m(m-1)\chi_2(Q_1\|Q_0)^2.
\end{align*}
Therefore,
\begin{align*}
\mathbb E B_m^4
=
\frac{3\chi_2(Q_1\|Q_0)^2}{m^2}
+
\frac{
\mathbb E_{Q_0}[A(Z)^4]
-3\chi_2(Q_1\|Q_0)^2
}{m^3}
=
O(m^{-2}).
\end{align*}

The absolute third moment can be bounded using the fourth moment.
By using the Lyapunov's inequality 
$\left(
\mathbb E |B_m|^3
\right)^{1/3}
\leq
\left(
\mathbb E |B_m|^4
\right)^{1/4},$ we have
\begin{align*}
\mathbb E |B_m|^3
\leq
\left(
\mathbb E [B_m^4]
\right)^{3/4}
=
O(m^{-3/2}).
\end{align*}
Note that although the absolute third moments 
$\mathbb E|B_m|^3=O(m^{-3/2}),$
the signed third moment satisfies the sharper estimate $
\mathbb E B_m^3=O(m^{-2}).$

\subsubsection{A Uniform Expansion of \(M_m(s)\)}
The purpose of this step is to identify the term of order \(m^{-1}\)
in
\begin{align*}
M_m(s)
=
\mathbb E_{Q_0^{\otimes m}}\left[(1+B_m)^s\right]
\end{align*}
and to bound the remaining terms uniformly by
\(c_{B,6,d}(1+|s|)^4m^{-2}\).  To do so, we expand \((1+B_m)^s\) through the cubic term. The cubic contribution is of order \(m^{-2}\) because
\(\mathbb E B_m^3=O(m^{-2})\), while the Taylor remainder is also of
order \(m^{-2}\) because it is controlled by
\(\mathbb E B_m^4=O(m^{-2})\). 

The Taylor expansion cannot be applied uniformly when \(B_m\) is
close to \(-1\), because the derivatives of \(x\mapsto(1+x)^s\) may
become large. We therefore use the event
$\mathcal E_m=\left\{|B_m|\leq\frac12\right\}$ introduced in
Section~\ref{sec:variational}, on which \(1+B_m\) is bounded away from zero. 
Thus, the Taylor expansion will only be used on \(\mathcal E_m\), whereas
the contribution from \(\mathcal E_m^{\mathrm c}\) will be controlled
directly using its exponentially small probability.

For fixed \(s\in\mathbb C\), consider the complex-valued function of
the real variable \(x\)
\begin{align*}
f_s(x)
\triangleq
(1+x)^s
=
\exp\bigl(s\log(1+x)\bigr),
\qquad x>-1.
\end{align*}
 Denote its cubic
Taylor polynomial at the origin by
\begin{align*}
T_3(s,x)
\triangleq
1+sx+\binom{s}{2}x^2+\binom{s}{3}x^3.
\end{align*}
and we also define the difference
\begin{align*}
R_4(s,x)
\triangleq
(1+x)^s-T_3(s,x).
\end{align*}

We now separate \(M_m(s)\) into three terms. The first term is the expectation of the cubic Taylor
polynomial and produces the leading term of the expansion. The
second is the Taylor remainder on \(\mathcal E_m\), which will be
controlled by \(\mathbb E B_m^4\). The third is the correction from
the exceptional event \(\mathcal E_m^{\mathrm c}\), which will be
exponentially small. More precisely,
\begin{align*}
M_m(s)
={}&
\mathbb E_{Q_0^{\otimes m}}\left[T_3(s,B_m)\right]+
\mathbb E_{Q_0^{\otimes m}}
\left[
R_4(s,B_m)\mathbf 1_{\mathcal E_m}
\right]
+
\mathbb E_{Q_0^{\otimes m}}
\left[
\bigl((1+B_m)^s-T_3(s,B_m)\bigr)
\mathbf 1_{\mathcal E_m^{\mathrm c}}
\right].
\end{align*}

We first evaluate the polynomial term, which determines the main
part of \(M_m(s)\). 
Note that
\begin{align*}
\mathbb E_{Q_0^{\otimes m}}\left[T_3(s,B_m)\right]
={}&
1+s\mathbb E_{Q_0^{\otimes m}}[B_m]
+\binom{s}{2}
\mathbb E_{Q_0^{\otimes m}}[B_m^2] +
\binom{s}{3}
\mathbb E_{Q_0^{\otimes m}}[B_m^3]\\
={}&
1+
\frac{\chi_2(Q_1\|Q_0)}{2m}s(s-1)
+
\binom{s}{3}
\frac{\mathbb E_{Q_0}[A(Z)^3]}{m^2}.
\end{align*}
Since, for a numerical constant \(c_{B,2}>0\),
$|\binom{s}{3}|
\leq
c_{B,2}(1+|s|)^3,$ we have
\begin{align*}
\left|
\binom{s}{3}
\frac{\mathbb E_{Q_0}[A(Z)^3]}{m^2}
\right|
\leq
\frac{c_{B,2}(1+|s|)^3}{m^2}.
\end{align*}
Thus,
\begin{align*}
\mathbb E_{Q_0^{\otimes m}}\left[T_3(s,B_m)\right]
=
1+
\frac{\chi_2(Q_1\|Q_0)}{2m}s(s-1)
+
O\left(\frac{(1+|s|)^3}{m^2}\right).
\end{align*}

We next control the Taylor remainder on \(\mathcal E_m\). Note that  Taylor's theorem with an integral remainder gives
\begin{align*}
R_4(s,x)
=
\frac{x^4}{3!}
\int_0^1
(1-t)^3f_s^{(4)}(tx)\,dt, \quad \text{where } f_s^{(4)}(x)
=
s(s-1)(s-2)(s-3)(1+x)^{s-4}.
\end{align*}
Suppose that
$\frac14
\leq
\operatorname{Re}(s)
\leq
\frac34$  and $
|x|\leq\frac12$.
For every \(t\in[0,1]\), we then have $
\frac12\leq1+tx\leq\frac32.$
Since \(1+tx\) is positive, we have
\begin{align*}
\left|(1+tx)^{s-4}\right|
=
(1+tx)^{\operatorname{Re}(s)-4}
\leq
2^{15/4}.
\end{align*}
Furthermore, for a numerical constant \(c_{B,3}>0\),
$|s(s-1)(s-2)(s-3)|
\leq
c_{B,3}(1+|s|)^4.$
Therefore, we have
\begin{align*}
|R_4(s,x)|
\leq
c_{B,3}(1+|s|)^4|x|^4,
\qquad |x|\leq\frac12.
\end{align*}
The
pointwise remainder bound and the fourth-moment estimate from Step~1
give
\begin{align*}
\left|
\mathbb E_{Q_0^{\otimes m}}
\left[
R_4(s,B_m)\mathbf 1_{\mathcal E_m}
\right]
\right|
&\leq
c_{B,3}(1+|s|)^4
\mathbb E_{Q_0^{\otimes m}}
\left[
B_m^4\mathbf 1_{\mathcal E_m}
\right]\\
&\leq
c_{B,3}(1+|s|)^4
\mathbb E_{Q_0^{\otimes m}}[B_m^4]\\
&\leq
\frac{c_{B,3}(1+|s|)^4}{m^2}.
\end{align*}

Finally, it remains to control the exceptional event. By the definitions of
\(a_{\min}\) and \(a_{\max}\),
\begin{align*}
a_{\min}\leq B_m\leq a_{\max}
\qquad
Q_0^{\otimes m}\text{-almost surely}.
\end{align*}
Moreover,
$1+B_m
=
\frac1m\sum_{j=1}^m
\frac{Q_1(Z_j)}{Q_0(Z_j)}
\geq0$
and \(1+B_m\leq1+a_{\max}\). Hence, whenever
\(1/4\leq\operatorname{Re}(s)\leq3/4\), there exists a constant
\(c_{B,4}>0\) such that
\begin{align*}
|(1+B_m)^s|
=
(1+B_m)^{\operatorname{Re}(s)}
\leq c_{B,4},
\end{align*}
where \(0^s\) is defined as zero because
\(\operatorname{Re}(s)>0\). The uniform boundedness of \(B_m\) also
implies, for a constant \(c_{B,5}>0\),
\begin{align*}
|T_3(s,B_m)|
\leq
c_{B,5}(1+|s|)^3.
\end{align*}
Therefore,
\begin{align*}
&\left|
\mathbb E_{Q_0^{\otimes m}}
\left[
\bigl((1+B_m)^s-T_3(s,B_m)\bigr)
\mathbf 1_{\mathcal E_m^{\mathrm c}}
\right]
\right|\\
&\quad\leq
c_{B,5}(1+|s|)^3
Q_0^{\otimes m}(\mathcal E_m^{\mathrm c})\\
&\quad\leq
c_{B,5}(1+|s|)^3 \times e^{-(1/(2(a_{\max}-a_{\min})^2))m}.
\end{align*}
Since \(e^{-(1/(2(a_{\max}-a_{\min})^2))m}\leq m^{-2}\) for all sufficiently large \(m\), this
term satisfies
\begin{align*}
\left|
\mathbb E_{Q_0^{\otimes m}}
\left[
\bigl((1+B_m)^s-T_3(s,B_m)\bigr)
\mathbf 1_{\mathcal E_m^{\mathrm c}}
\right]
\right|
\leq
\frac{c_{B,5}(1+|s|)^4}{m^2}.
\end{align*}
Combining the three terms above yields
\begin{align}
M_m(s)
=
1+
\frac{\chi_2(Q_1\|Q_0)}{2m}s(s-1)
+
R_m(s), \label{eq:0915}
\end{align}
where, for every fixed \(d>0\), there exists \(c_{B,6,d}>0\) such that
$|R_m(s)|
\leq
\frac{c_{B,6,d}(1+|s|)^4}{m^2}$
uniformly over all $s$ such that
$\frac14
\leq
\operatorname{Re}(s)
\leq
\frac34,
|\operatorname{Im}(s)|
\leq
d\sqrt{\log m}.$
\medskip
\subsubsection{Passing From \(M_m(s)\) To \(\log M_m(s)\)}
Applying the triangle inequality to the expansion in~\eqref{eq:0915} yields
\begin{align*}
|M_m(s)-1|
&\le
\frac{\chi_2(Q_1\|Q_0)}{2m}|s(s-1)|
+
|R_m(s)| \\
&\leq \frac{\chi_2(Q_1\|Q_0)}{2}\frac{(1+|s|)^2}{m}
+
c_{B,6,d}\frac{(1+|s|)^4}{m^2},
\end{align*}
where the last inequality follows from
$|s(s-1)|
=
|s|\,|s-1| \le |s|(|s|+1) \le (1+|s|)^2.$

On the range
$\frac14\le\operatorname{Re}(s)\le\frac34$ and 
$|\operatorname{Im}(s)|\le d\sqrt{\log m}$, we have
\begin{align*}
|s|
&\le
|\operatorname{Re}(s)|
+
|\operatorname{Im}(s)| \le
\frac34+d\sqrt{\log m}.
\end{align*}
For \(m\ge e\), we have \(\sqrt{\log m}\ge1\) and hence
\begin{align*}
1+|s|
\le
\frac74+d\sqrt{\log m}
\le
\left(d+\frac74\right)\sqrt{\log m}.
\end{align*}
Therefore,
\begin{align*}
|M_m(s)-1|
&\le
\frac{\chi_2(Q_1\|Q_0)}{2} \left(d+\frac74\right)^2\frac{\log m}{m} +
c_{B,6,d}\left(d+\frac74\right)^4
\frac{(\log m)^2}{m^2}.
\end{align*}
Both terms converge to zero as \(m\to\infty\), which holds for all
\(s\) satisfying the condition of Lemma 2.
In particular, for all sufficiently large
\(m\),
$|M_m(s)-1|\le\frac12.$
We  define \(\log M_m(s)\) by the absolutely convergent power
series
\begin{align*}
\log M_m(s)
=
\sum_{k=1}^{\infty}
\frac{(-1)^{k+1}}{k}
\bigl(M_m(s)-1\bigr)^k.
\end{align*}
This series defines the unique analytic logarithm in a neighborhood
of one whose value at one is zero. Thus, although \(M_m(s)\) may be
complex, the quantity \(\log M_m(s)\) is unambiguously defined.

Separating the first-order term in  this series gives
\begin{align*}
\log M_m(s)
=
M_m(s)-1
+
\sum_{k=2}^{\infty}
\frac{(-1)^{k+1}}{k}
\bigl(M_m(s)-1\bigr)^k.
\end{align*}
Since \(|M_m(s)-1|\le1/2\), we have
\begin{align*}
\left|
\log M_m(s)-\bigl(M_m(s)-1\bigr)
\right|
&\le
\sum_{k=2}^{\infty}|M_m(s)-1|^k =
\frac{|M_m(s)-1|^2}
{1-|M_m(s)-1|}\le
2|M_m(s)-1|^2.
\end{align*}
Substituting the expansion obtained in~\eqref{eq:0915} now yields
\begin{align*}
\log M_m(s)
={}&
M_m(s)-1
+
\left[
\log M_m(s)-\bigl(M_m(s)-1\bigr)
\right]\\
={}&
\frac{\chi_2(Q_1\|Q_0)}{2m}s(s-1)
+
R_m(s)+
\left[
\log M_m(s)-\bigl(M_m(s)-1\bigr)
\right].
\end{align*}
Accordingly, define
\begin{align*}
r_m(s)
\triangleq
R_m(s)
+
\left[
\log M_m(s)-\bigl(M_m(s)-1\bigr)
\right],
\end{align*}
which absorbs the Taylor-expansion remainder
\(R_m(s)\) and the imprecision  caused by taking
the logarithm. Hence,
\begin{align*}
|r_m(s)|
&\le
|R_m(s)|
+
\left|
\log M_m(s)-\bigl(M_m(s)-1\bigr)
\right|\\
&\le
\frac{c_{B,6,d}(1+|s|)^4}{m^2} + 2|M_m(s)-1|^2 \\
&\le \frac{c_{B,1,d}(1+|s|)^4}{m^2}
\end{align*}
for some constant $c_{B,1,d}$.
We have therefore proved that
\begin{align*}
\log M_m(s)
=
\frac{\chi_2(Q_1\|Q_0)}{2m}s(s-1)
+
r_m(s),
\qquad
|r_m(s)|
\le
\frac{c_{B,1,d}(1+|s|)^4}{m^2}.
\end{align*}
This completes the proof of Lemma~2.
\hfill
\subsection{Proof of Proposition 1}
Recall that $L_{m,\ell} = \sum_{j=1}^{\ell} \log \frac{P_m(\mathbf Z_j)}
{Q_0^{\otimes m}(\mathbf Z_j)}$. By the definitions of $\rho_m$ and $H_m$ in~\eqref{rho}, we have 
\begin{align*}
\rho_m = M_m(1/2) \quad \text{and} \quad H_m(\mathbf z)
=
\frac{
Q_0^{\otimes m}(\mathbf z)
}{
M_m(1/2)
}
\left(
\frac{P_m(\mathbf z)}
{Q_0^{\otimes m}(\mathbf z)}
\right)^{1/2}.
\end{align*}
Therefore, the characteristic function of the log-likelihood ratio of one PPM block under \(H_m\) is
\begin{align*}
\mathbb E_{H_m}
\left[
\exp\left(
it\log
\frac{P_m(\mathbf Z_1)}
{Q_0^{\otimes m}(\mathbf Z_1)}
\right)
\right]
=
\frac{1}{M_m(1/2)}
\sum_{\mathbf z}
Q_0^{\otimes m}(\mathbf z)
\left(
\frac{P_m(\mathbf z)}
{Q_0^{\otimes m}(\mathbf z)}
\right)^{1/2+it} =
\frac{M_m(1/2+it)}
{M_m(1/2)}.
\end{align*}
Consequently, independence across the \(\ell\) blocks yields
\begin{align}
\varphi_{L_{m,\ell}}(t)
=
\mathbb E_{H_m^{\otimes\ell}}
\left[e^{itL_{m,\ell}}\right] =
\left[
\frac{M_m(1/2+it)}
{M_m(1/2)}
\right]^\ell .
\label{eq:ell-block-cf}
\end{align}

Note that
\begin{align*}
\log M_m(1/2+it)
=
-\frac{\chi_2(Q_1\|Q_0)}{8m}
-\frac{\chi_2(Q_1\|Q_0)}{2m}t^2
+r_m(1/2+it),
\end{align*}
\begin{align*}
\log M_m(1/2)
=
-\frac{\chi_2(Q_1\|Q_0)}{8m}
+r_m(1/2).
\end{align*}
Substitution into the expression for
\(\varphi_{L_{m,\ell}}(t)\) in \eqref{eq:ell-block-cf} yields
\begin{align}
\varphi_{L_{m,\ell}}(t)
=
\exp\left\{
-\frac{\ell\chi_2(Q_1\|Q_0)}{2m}t^2
+\ell\left[r_m(1/2+it)-r_m(1/2)\right]
\right\}.
\label{eq:cf-expansion}
\end{align}
It then follows from \eqref{eq:cf-expansion} that
\begin{align*}
&\left|
\varphi_{L_{m,\ell}}(t)
-
e^{-\ell\chi_2(Q_1\|Q_0)t^2/(2m)}
\right| \\
&=
e^{-\ell\chi_2(Q_1\|Q_0)t^2/(2m)}
\times \left|
 \exp\left\{
\ell\left[
r_m(1/2+it)-r_m(1/2)
\right]
\right\}
-1
\right| \\
&\leq
e^{-\ell\chi_2(Q_1\|Q_0)t^2/(2m)}
\times \left|
\ell\left[
r_m(1/2+it)-r_m(1/2)
\right]
\right| 
\times
\exp\left\{
\left|
\ell\left[
r_m(1/2+it)-r_m(1/2)
\right]
\right|
\right\},
\end{align*}
where the last inequality follows from
the fact that \(|e^w-1|\leq |w|e^{|w|}\) for \(w\in\mathbb C\).

On the  range $|t|\leq d\sqrt{\log m} $, it follows from Lemma 2 that 
\begin{align*}
\left|
\ell[r_m(1/2+it)-r_m(1/2)]
\right|
\le
\frac{c_{B,1,d}(1+|t|^4)}m 
\le \frac{c_{B,1,d}[1+d^4(\log m)^2]}{m} 
=o(1),
\end{align*}
and hence the  exponential of this term is also bounded
by a constant for all sufficiently large \(m\). Moreover, since
\(\ell/m\geq c_0\) due to the condition of Theorem~1, we have
\begin{align*}
e^{-\ell\chi_2(Q_1\|Q_0)t^2/(2m)}
\leq e^{-c_{B,7}t^2},
\qquad \text{where }
c_{B,7}\triangleq
\frac{c_0\chi_2(Q_1\|Q_0)}{2}>0.
\end{align*}
Therefore,
\begin{align}
\left|
\varphi_{L_{m,\ell}}(t)
-
e^{-\ell\chi_2(Q_1\|Q_0)t^2/(2m)}
\right|
\leq
\frac{c_{B,1,d}}{m}(1+|t|^4) \times e^{-c_{B,7}t^2}. \label{phi}
\end{align}
Since \(e^{-t^2/(2m)}\leq1\), the bound on the difference between the two characteristic functions in~\eqref{phi} yields 
\begin{align*}
\int_{|t|\leq d\sqrt{\log m}}
\mathcal E_{m,\ell}(t)\,dt
&\leq
\frac{c_{B,1,d}}{m}
\int_{|t|\leq d\sqrt{\log m}}
\frac{(1+|t|^4)e^{-c_{B,7}t^2}}{t^2+1/4}\,dt\\
&\leq
\frac{c_{B,1,d}}{m}
\int_{\mathbb R}
\frac{(1+|t|^4)e^{-c_{B,7}t^2}}{t^2+1/4}\,dt\\
&=
O(m^{-1}),
\end{align*}
where the last step follows from the fact that  the integral over $\mathbb{R}$ is finite and independent of \(m\). This completes the proof of Proposition 1.

\section{Proof of Proposition 2} \label{appendix:proposition2}

\par\noindent{\bfseries Proposition 2.} There exist \(u_0>0\) and \(d_0>0\) such that, for every fixed \(d\ge d_0\),
\begin{align*}
\int_{d\sqrt{\log m}<|t|\le u_0\sqrt m}
\mathcal E_{m,\ell}(t)\,dt
=O(m^{-1}).
\end{align*}

\subsection{Proof of An Auxiliary Result}
In this subsection, we first prove that there exist constants $u_0 > 0$ and $c_{C,1} > 0$  such that 
\begin{align*}
\left|
\mathbb E_{H_m}
\exp\left(
iu\sqrt m
\log\frac{P_m(\mathbf Z)}
{Q_0^{\otimes m}(\mathbf Z)}
\right)
\right|
\leq
e^{-c_{C,1}u^2},
\qquad |u|\leq u_0.
\end{align*}

For simplicity, we define
\begin{align}
X_m
\triangleq
\sqrt m
\log
\frac{P_m(\mathbf Z)}
{Q_0^{\otimes m}(\mathbf Z)}, \quad \text{where} \ \mathbf Z\sim H_m.  \label{eq:xm}
\end{align}
Let \(X_m'\) be an independent copy of \(X_m\). By independence, we have
\begin{align*}
\left|
\mathbb E_{H_m}e^{iuX_m}
\right|^2
&= \mathbb E_{H_m}e^{iuX_m}
\times
\overline{
\mathbb E_{H_m}e^{iuX_m}
} =
\mathbb E_{H_m}e^{iuX_m} \times 
\mathbb E_{H_m}e^{-iuX_m'} =
\mathbb E e^{iu(X_m-X_m')}.
\end{align*}
Since \(X_m\) and \(X_m'\) are independent and identically
distributed, the pairs \((X_m,X_m')\) and \((X_m',X_m)\) have the
same distribution. Consequently,
\begin{align*}
X_m-X_m'
\overset{\mathrm d}=
X_m'-X_m
=
-(X_m-X_m').
\end{align*}
Thus the distribution of \(X_m-X_m'\) is symmetric about zero.

Since the sine function is odd, this symmetry implies $
\mathbb E\sin\bigl(u(X_m-X_m')\bigr)=0.$
Therefore,
\begin{align*}
\mathbb E e^{iu(X_m-X_m')}
&=
\mathbb E\cos\bigl(u(X_m-X_m')\bigr) +
i\mathbb E\sin\bigl(u(X_m-X_m')\bigr) =
\mathbb E\cos\bigl(u(X_m-X_m')\bigr).
\end{align*}
It follows that
\begin{align*}
1-
\left|
\mathbb E_{H_m}e^{iuX_m}
\right|^2
=
\mathbb E\left[
1-\cos\bigl(u(X_m-X_m')\bigr)
\right].
\end{align*}
This is useful because the elementary inequality 
$1-\cos y
\ge
\frac{y^2}{2}
-
\frac{|y|^3}{6}$ for 
$y\in\mathbb R$
converts the problem of estimating characteristic function into estimating the
second and third moments of \(X_m-X_m'\).

We first evaluate these moments. In Appendix~\ref{app:midpoint-moments}, we show that
\begin{align*}
&\operatorname{Var}_{H_m}(X_m)
=
\chi_2(Q_1\|Q_0)+O(m^{-1}), \qquad  \mathbb E_{H_m}
\left|
X_m-\mathbb E_{H_m}X_m
\right|^3 =
O(1).
\end{align*}
Therefore,
\begin{align*}
\mathbb E(X_m-X_m')^2
&=
2\operatorname{Var}_{H_m}(X_m) =
2\chi_2(Q_1\|Q_0)+O(m^{-1}).
\end{align*}
 Consequently, for all sufficiently large
\(m\),
\begin{align*}
\mathbb E(X_m-X_m')^2
\ge
\chi_2(Q_1\|Q_0).
\end{align*}
Furthermore, note that 
\begin{align*}
X_m-X_m'
=
\bigl(X_m-\mathbb E_{H_m}X_m\bigr)
-
\bigl(X_m'-\mathbb E_{H_m}X_m'\bigr).
\end{align*}
Using
$|x-y|^3\le4(|x|^3+|y|^3),$
we obtain
\begin{align*}
\mathbb E|X_m-X_m'|^3
&\le
4\mathbb E
\left|
X_m-\mathbb E_{H_m}X_m
\right|^3 +
4\mathbb E
\left|
X_m'-\mathbb E_{H_m}X_m'
\right|^3 =
O(1).
\end{align*}
Thus there exists a constant \(c_{C,2}>0\) such that for sufficiently large $m$,
\begin{align*}
\mathbb E|X_m-X_m'|^3\le c_{C,2},
\end{align*}

We now apply the cosine inequality. For sufficiently large $m$
\begin{align*}
1-
\left|
\mathbb E_{H_m}e^{iuX_m}
\right|^2
&=
\mathbb E\left[
1-\cos\bigl(u(X_m-X_m')\bigr)
\right]\\
&\ge
\frac{u^2}{2}
\mathbb E(X_m-X_m')^2
-
\frac{|u|^3}{6}
\mathbb E|X_m-X_m'|^3\\
&\ge
\frac{\chi_2(Q_1\|Q_0)}{2}u^2
-
\frac{c_{C,2}}{6}|u|^3.
\end{align*}
To ensure that the quadratic term
dominates this error, choose \(u_0>0\) sufficiently small that
$\frac{c_{C,2}}{6}u_0
\le
\frac{\chi_2(Q_1\|Q_0)}{4}.$
Then, for every \(|u|\le u_0\),
$\frac{c_{C,2}}{6}|u|^3
\le
\frac{\chi_2(Q_1\|Q_0)}{4}u^2,$
and therefore
\begin{align*}
1-
\left|
\mathbb E_{H_m}e^{iuX_m}
\right|^2
\ge
\frac{\chi_2(Q_1\|Q_0)}{4}u^2.
\end{align*}
Equivalently,
\begin{align*}
\left|
\mathbb E_{H_m}e^{iuX_m}
\right|^2
\le
1-
\frac{\chi_2(Q_1\|Q_0)}{4}u^2.
\end{align*}
We can further choose \(u_0\) such that 
$0\le
\frac{\chi_2(Q_1\|Q_0)}{4}u^2
\le1$ for all $|u|\le u_0.$
Using \(1-x\le e^{-x}\) for \(x\ge0\), we obtain
\begin{align*}
\left|
\mathbb E_{H_m}e^{iuX_m}
\right|^2
\le
\exp\left(
-\frac{\chi_2(Q_1\|Q_0)}{4}u^2
\right).
\end{align*}
Therefore, we obtain
\begin{align*}
\left|
\mathbb E_{H_m}
\exp\left(
iu\sqrt m
\log
\frac{P_m(\mathbf Z)}
{Q_0^{\otimes m}(\mathbf Z)}
\right)
\right|
\le
\exp\left(
-c_{C,1} u^2
\right),
\qquad |u|\le u_0.
\end{align*}
where $c_{C,1}=\frac{\chi_2(Q_1\|Q_0)}{8}>0$.

\subsection{Proof of Proposition~2}
For
\(|t|\leq u_0\sqrt m\), taking \(u=t/\sqrt m\) yields
\begin{align*}
\left|
\mathbb E_{H_m}
\exp\left(
it\log\frac{P_m(\mathbf Z)}
{Q_0^{\otimes m}(\mathbf Z)}
\right)
\right|
\leq
e^{-c_{C,1}t^2/m}.
\end{align*}
Since \(\varphi_{L_{m,\ell}}(t)\) is the \(\ell\)-th power of this
characteristic function, we have
\begin{align}
|\varphi_{L_{m,\ell}}(t)|
\leq
e^{-c_{C,1}\ell t^2/m}
\leq
e^{-c_{C,1}c_0t^2},
\label{eq:quadratic-decay}
\end{align}
where the last inequality follows from \(\ell/m\geq c_0\).
The characteristic function of Gaussian variable $N_{m,\ell}$ satisfies
\begin{align*}
e^{-\ell\chi_2(Q_1\|Q_0)t^2/(2m)}
\leq
e^{-c_0\chi_2(Q_1\|Q_0)t^2/2}.
\end{align*}
Define
$b
\triangleq
\min\left\{
c_{C,1}c_0,
\frac{c_0\chi_2(Q_1\|Q_0)}{2}
\right\}>0.$
The triangle inequality and \eqref{eq:quadratic-decay} then give
\begin{align*}
&
\left|
\varphi_{L_{m,\ell}}(t)
-
e^{-\ell\chi_2(Q_1\|Q_0)t^2/(2m)}
\right| \leq
|\varphi_{L_{m,\ell}}(t)|
+
e^{-\ell\chi_2(Q_1\|Q_0)t^2/(2m)}
\leq
2e^{-bt^2}.
\end{align*}
Recalling the definition
\begin{align*}
\mathcal E_{m,\ell}(t)
=
\frac{e^{-t^2/(2m)}}{t^2+1/4}
\left|
\varphi_{L_{m,\ell}}(t)
-
e^{-\ell\chi_2(Q_1\|Q_0)t^2/(2m)}
\right|,
\end{align*}
and using the facts  $e^{-t^2/(2m)}\leq1$ and  $\frac{1}{t^2+1/4}\leq4$,
we obtain
$\mathcal E_{m,\ell}(t)
\leq
8e^{-bt^2}$
throughout the range considered in Proposition~2. Consequently,
\begin{align*}
&
\int_{d\sqrt{\log m}<|t|\leq u_0\sqrt m}
\mathcal E_{m,\ell}(t)\,dt
\leq
8\int_{|t|>d\sqrt{\log m}}
e^{-bt^2}\,dt.
\end{align*}
For every \(x>0\),
\begin{align*}
\int_{|t|>x}e^{-bt^2}\,dt
\leq
e^{-bx^2/2}
\int_{\mathbb R}e^{-bt^2/2}\,dt
=
\sqrt{\frac{2\pi}{b}}\,
e^{-bx^2/2}.
\end{align*}
Taking \(x=d\sqrt{\log m}\) therefore gives
\begin{align*}
\int_{d\sqrt{\log m}<|t|\leq u_0\sqrt m}
\mathcal E_{m,\ell}(t)\,dt
&\leq
8\sqrt{\frac{2\pi}{b}}\,
e^{-bd^2\log m/2}
=
8\sqrt{\frac{2\pi}{b}}\,
m^{-bd^2/2}.
\end{align*}
Choose \(d_0>0\) such that
$\frac{bd_0^2}{2}\geq1$.
Then, for every \(d\geq d_0\), we have
$m^{-bd^2/2}\leq m^{-1}.
$
This completes the proof of Proposition 2. \qed

\section{Proof of Proposition 3} \label{appendix:proposition3}
\par\noindent{\bfseries Proposition 3.}
For every fixed \(k>0\), there exists a constant \(c_{3,k}>0\) such that
\begin{align*}
\int_{u_0\sqrt m<|t|\leq\sqrt{km\log m}}
\mathcal E_{m,\ell}(t)\,dt
=
O(e^{-c_{3,k}m}),
\end{align*}
where \(u_0>0\) is the constant chosen in Proposition 2.

\subsection{Proof of An Auxiliary Result}

In this subsection, we first prove that for every fixed \(k>0\), there exists a
constant \(\eta_k\in(0,1)\) such that
\begin{align*}
\left|\mathbb E_{H_m}e^{iuX_m} \right| = \left|
\mathbb E_{H_m}
\exp\left(
iu\sqrt m
\log\frac{P_m(\mathbf Z)}
{Q_0^{\otimes m}(\mathbf Z)}
\right)
\right|
\leq
\eta_k
\end{align*}
uniformly over $u_0\leq |u|\leq\sqrt{k\log m}$, where the random variable \(X_m\) is introduced in Appendix~\ref{appendix:proposition2}.
Note that random variable \(X_m\) is not a sum of independent random variables under \(H_m\). We therefore introduce
\begin{align*}
Y_m
\triangleq
\sqrt m\,B_m
=
\frac1{\sqrt m}\sum_{j=1}^m A(Z_j),
\end{align*}
where \(A(.)\) is the single-letter quantity introduced in Appendix~\ref{appendix:proposition1}.
Under \(Q_0^{\otimes m}\), the characteristic function of \(Y_m\)
factors into \(m\) single-letter characteristic functions. The proof
will show that replacing \(X_m\) by \(Y_m\), and then replacing
\(H_m\) by \(Q_0^{\otimes m}\), introduces only a vanishing
error.

\medskip
\subsubsection{Replacing \(X_m\) By \(Y_m\) under \(H_m\)}
On the event \(\mathcal E_m\) introduced in Section~\ref{sec:variational}, 
the quantity \(1+B_m\) is bounded below by \(1/2\). Using
$\log(1+x)-x
=
-\int_0^x\frac{t}{1+t}\,dt$ for
$x>-1$,
we obtain that, for \(|x|\le1/2\),
\begin{align*}
|\log(1+x)-x|
&\le
\int_0^{|x|}
\frac{t}{1-t}\,dt \le
\int_0^{|x|}2t\,dt =
|x|^2.
\end{align*}
Applying this inequality with \(x=B_m\) leads to
\begin{align*}
|\log(1+B_m)-B_m|
\le
B_m^2
\end{align*}
on \(\mathcal E_m\). Therefore, we have
\begin{align*}
|X_m-Y_m|
=
\sqrt m\,|\log(1+B_m)-B_m|
\le
\sqrt m\,B_m^2.
\end{align*}

We next take the expectation under \(H_m\). First, one can verify that 
\begin{align*}
\frac{H_m(\mathbf z)}{Q_0^{\otimes m}(\mathbf z)}
=
\frac{\sqrt{1+B_m(\mathbf z)}}{\rho_m}.
\end{align*}
Since \(A(Z)\) is bounded, \(B_m\) is bounded uniformly in \(m\), and
hence \(\sqrt{1+B_m}\) is also uniformly bounded. Moreover,
\begin{align*}
|\rho_m-1|
&=
\left|
\mathbb E_{Q_0^{\otimes m}}\sqrt{1+B_m}-1
\right| \le
\mathbb E_{Q_0^{\otimes m}}
\left|\sqrt{1+B_m}-1\right| \le
\mathbb E_{Q_0^{\otimes m}}|B_m| \le
\sqrt{\mathbb E_{Q_0^{\otimes m}}B_m^2} =
\sqrt{\frac{\chi_2(Q_1\|Q_0)}{m}}.
\end{align*}
Thus, for all sufficiently large \(m\), we have \(\rho_m\ge1/2\), and hence there exists a constant $c_{D,1}>0$ such that for all~$\mathbf{z}$,
\begin{align*}
H_m(\mathbf z)
&\le
c_{D,1}\,Q_0^{\otimes m}(\mathbf z).
\end{align*}
Therefore,
\begin{align*}
\mathbb E_{H_m}
\left[
|X_m-Y_m|
\mathbf 1_{\{|B_m|\le1/2\}}
\right]
&\le
\sqrt m\,
\mathbb E_{H_m}
\left[
B_m^2\mathbf 1_{\{|B_m|\le1/2\}}
\right]\\
&\le
c_{D,1}\sqrt m\,
\mathbb E_{Q_0^{\otimes m}}B_m^2\\
&=
c_{D,1}\sqrt m\,
\frac{\chi_2(Q_1\|Q_0)}{m} =
O(m^{-1/2}).
\end{align*}

It remains to control the event \(\{|B_m|>1/2\}\). Since \eqref{eq:Bm-tail} shows that 
$Q_0^{\otimes m}\!\left(|B_m|>\frac12\right)
\le
2\exp\left(
-\frac{m}{2(a_{\max}-a_{\min})^2}
\right),$
we obtain
\begin{align*}
H_m\{|B_m|>1/2\}
\le 2c_{D,1}\exp\left(
-\frac{m}{2(a_{\max}-a_{\min})^2}
\right).
\end{align*}
By calculating that $\sup_{\mathcal{E}_m^c \cap \text{supp}(H_m)} |X_m - Y_m| = O(\sqrt{m}\log m)$, we obtain
\begin{align*}
\mathbb E_{H_m}
\left[
|X_m-Y_m|
\mathbf 1_{\{|B_m|>1/2\}}
\right] =
O(e^{-(1/2(a_{\max}-a_{\min})^2)m}).
\end{align*}
Combining the two events together yields that
\begin{align*}
\mathbb E_{H_m}|X_m-Y_m|
=
O(m^{-1/2}).
\end{align*}
This further helps to control the change in the characteristic function
caused by replacing \(X_m\) with \(Y_m\). Since
$|e^{ia}-e^{ib}|\le |a-b|$ for  $a,b\in\mathbb R$,
we have, for some constant \(c_{D,2}>0\),
\begin{align*}
\left|
\mathbb E_{H_m}e^{iuX_m}
-
\mathbb E_{H_m}e^{iuY_m}
\right|
&\le
\mathbb E_{H_m}
\left|
e^{iuX_m}-e^{iuY_m}
\right| \le
|u|\mathbb E_{H_m}|X_m-Y_m| \le
\frac{c_{D,2}|u|}{\sqrt m}.
\end{align*}

\medskip
\subsubsection{Replacing \(H_m\) By \(Q_0^{\otimes m}\)}
We next show that the expectation of \(e^{iuY_m}\) changes by only
\(O(m^{-1/2})\) when the underlying distribution is changed from
\(H_m\) to \(Q_0^{\otimes m}\). By the definition of \(H_m\),
\begin{align*}
\sum_{\mathbf z}
\left|
H_m(\mathbf z)-Q_0^{\otimes m}(\mathbf z)
\right|
&=
\mathbb E_{Q_0^{\otimes m}}
\left|
\frac{\sqrt{1+B_m}}{\rho_m}-1
\right| =
\frac1{\rho_m}
\mathbb E_{Q_0^{\otimes m}}
\left|
\sqrt{1+B_m}-\rho_m
\right|.
\end{align*}
Using the triangle inequality, we have
\begin{align*}
\left|\sqrt{1+B_m}-\rho_m\right|
\le
\left|\sqrt{1+B_m}-1\right|
+
|1-\rho_m|.
\end{align*}
Note that 
\begin{align*}
\mathbb E_{Q_0^{\otimes m}}
\left|\sqrt{1+B_m}-1\right|
\le \mathbb E_{Q_0^{\otimes m}} |B_m| \le \sqrt{\mathbb E_{Q_0^{\otimes m}}[B_m^2]} =
\sqrt{\frac{\chi_2(Q_1\|Q_0)}{m}}
\end{align*}
and
\begin{align*}
|1-\rho_m|
&=
\left|
\mathbb E_{Q_0^{\otimes m}}
\left[
1-\sqrt{1+B_m}
\right]
\right|\le
\mathbb E_{Q_0^{\otimes m}}
\left|
1-\sqrt{1+B_m}
\right| \le
\sqrt{\frac{\chi_2(Q_1\|Q_0)}{m}}.
\end{align*}
Since \(\rho_m\ge1/2\) for all sufficiently large \(m\), there exists a constant $c_{D,3}>0$ such that
\begin{align*}
\sum_{\mathbf z}
\left|
H_m(\mathbf z)-Q_0^{\otimes m}(\mathbf z)
\right|
\le
\frac{c_{D,3}}{\sqrt m}.
\end{align*}
Because \(|e^{iuY_m}|=1\), we have
\begin{align*}
\left|
\mathbb E_{H_m}e^{iuY_m}
-
\mathbb E_{Q_0^{\otimes m}}e^{iuY_m}
\right|
&\le
\sum_{\mathbf z}
\left|
H_m(\mathbf z)-Q_0^{\otimes m}(\mathbf z)
\right|
|e^{iuY_m(\mathbf z)}| \le
\frac{c_{D,3}}{\sqrt m}.
\end{align*}

Combining the above analyses, we obtain, for some constant \(c_{D,4}>0\),
\begin{align*}
\left|
\mathbb E_{H_m}e^{iuX_m}
-
\mathbb E_{Q_0^{\otimes m}}e^{iuY_m}
\right|
&\le
\left|
\mathbb E_{H_m}e^{iuX_m}
-
\mathbb E_{H_m}e^{iuY_m}
\right| +
\left|
\mathbb E_{H_m}e^{iuY_m}
-
\mathbb E_{Q_0^{\otimes m}}e^{iuY_m}
\right|\\
&\le
\frac{c_{D,4}(1+|u|)}{\sqrt m}.
\end{align*}
This allows us to replace the
nonlinear midpoint variable \(X_m\) by the sum of i.i.d. random variables \(Y_m\).

\medskip
\subsubsection{Calculating the Characteristic Function of
\(Y_m\)}
Define the single-letter characteristic function
\begin{align*}
\varphi_A(s)
\triangleq
\mathbb E_{Q_0}e^{isA(Z)},
\end{align*}
and note that
\begin{align*}
\mathbb E_{Q_0^{\otimes m}} [e^{iuY_m}]
&=
\mathbb E_{Q_0^{\otimes m}}
\exp\left(
\frac{iu}{\sqrt m}
\sum_{j=1}^m A(Z_j)
\right)=
\varphi_A(u/\sqrt m)^m.
\end{align*}

We next prove that \(|\varphi_A(s)|\) has a quadratic gap below one
for sufficiently small \(s\). Let \(A(Z')\) be an independent copy of
\(A(Z)\). Symmetrization implies
\begin{align*}
|\varphi_A(s)|^2
&=
\mathbb E
[e^{is(A(Z)-A(Z'))}] =
\mathbb E
[\cos\left(s(A(Z)-A(Z'))\right)],
\end{align*}
where the imaginary part vanishes because
\(A(Z)-A(Z')\) has a symmetric distribution. Hence
\begin{align*}
1-|\varphi_A(s)|^2
=
\mathbb E
\left[
1-\cos\left(s(A(Z)-A(Z'))\right)
\right].
\end{align*}
Using 
$1-\cos x
\ge
\frac{x^2}{2}-\frac{|x|^3}{6}$ for 
$x\in\mathbb R$,
we obtain
\begin{align*}
1-|\varphi_A(s)|^2
\ge{}&
\frac{s^2}{2}
\mathbb E\left[
(A(Z)-A(Z'))^2
\right] -
\frac{|s|^3}{6}
\mathbb E\left[
|A(Z)-A(Z')|^3
\right].
\end{align*}
Note that $
\mathbb E\left[
(A(Z)-A(Z'))^2
\right]
=
2\chi_2(Q_1\|Q_0).$
The third absolute moment is finite because \(A(Z)\) is bounded.
Therefore, there exists a constant $c_{D,5}>0$ such that 
\begin{align*}
1-|\varphi_A(s)|^2
\ge
\chi_2(Q_1\|Q_0)s^2-c_{D,5}|s|^3.
\end{align*}
Choosing \(s_0>0\) sufficiently small implies that for all $|s| \leq s_0$,
\begin{align*}
1-|\varphi_A(s)|^2
\ge
\frac{\chi_2(Q_1\|Q_0)}{2}s^2.
\end{align*}
Let $c_{D,6} \triangleq \frac{\chi_2(Q_1\|Q_0)}{4}$.
Hence
\begin{align*}
|\varphi_A(s)|
\le
\exp(-c_{D,6}s^2),
\qquad |s|\le s_0.
\end{align*}
Fix \(k>0\). If
$|u|\le\sqrt{k\log m},$
then
$\frac{|u|}{\sqrt m}
\le
\sqrt{\frac{k\log m}{m}},$
which is smaller than \(s_0\) for all sufficiently large \(m\).
Therefore,
\begin{align*}
\left|
\mathbb E_{Q_0^{\otimes m}}e^{iuY_m}
\right|
&=
|\varphi_A(u/\sqrt m)|^m\le
\left[
\exp\left(
-c_{D,6}\frac{u^2}{m}
\right)
\right]^m =
e^{-c_{D,6}u^2}.
\end{align*}
Combining this bound with the approximation obtained in Steps 1 and
2 gives
\begin{align*}
\left|\mathbb E_{H_m}e^{iuX_m} \right|
\le
e^{-c_{D,6}u^2}
+
\frac{c_{D,4}(1+|u|)}{\sqrt m},
\qquad
|u|\le\sqrt{k\log m}.
\end{align*}
Finally, letting
$\eta_k
\triangleq
\frac{1+e^{-c_{D,6}u_0^2}}{2} \in  (0,1)$, one can check that for all $u_0\le |u|\le\sqrt{k\log m}$, 
$\left|\mathbb E_{H_m}e^{iuX_m} \right| \le \eta_k$.

\subsection{Proof of Proposition 3}
Consider the range
$u_0\sqrt m<|t|\leq\sqrt{km\log m}$. Taking \(u=t/\sqrt m\) places \(u\) in the  range  $u_0<|u|\leq\sqrt{k\log m}$. By independence across the \(\ell\) blocks, we have
\begin{align*}
|\varphi_{L_{m,\ell}}(t)| \leq
\eta_k^\ell =
\exp\left(\ell\log\eta_k\right) 
&\leq
\exp\left(c_0m\log\eta_k\right)=
e^{-c_{D,7,k}m},
\end{align*}
where
$c_{D,7,k}\triangleq-c_0\log\eta_k>0.$

The characteristic function of Gaussian variable $N_{m,\ell}$ is exponentially small
on the same range. Indeed, using \(\ell/m\geq c_0\) and
\(|t|\geq u_0\sqrt m\), we obtain
\begin{align*}
e^{-\ell\chi_2(Q_1\|Q_0)t^2/(2m)}
\leq
e^{-c_0\chi_2(Q_1\|Q_0)t^2/2} \leq
e^{-c_{D,8}m},
\end{align*}
where
$c_{D,8}
\triangleq
\frac{c_0\chi_2(Q_1\|Q_0)u_0^2}{2}
>0.$
Let $c_{D,9,k}\triangleq\min\{c_{D,7,k},c_{D,8}\}>0.$ It follows from the triangle inequality that
\begin{align*}
\left|
\varphi_{L_{m,\ell}}(t)
-
e^{-\ell\chi_2(Q_1\|Q_0)t^2/(2m)}
\right|
\leq
2e^{-c_{D,9,k}m}.
\end{align*}
Since
$e^{-t^2/(2m)}\leq1$
and $\frac{1}{t^2+1/4}\leq4$,
the definition of \(\mathcal E_{m,\ell}(t)\) yields $
\mathcal E_{m,\ell}(t)
\leq
8e^{-c_{D,9,k}m}$
throughout the uniform-contraction range. The total length of this
range is at most
$2\sqrt{km\log m}.$
Therefore,
\begin{align*}
\int_{u_0\sqrt m<|t|\leq\sqrt{km\log m}}
\mathcal E_{m,\ell}(t)\,dt
\leq
16\sqrt{km\log m}\,e^{-c_{D,9,k}m}.
\end{align*}
Since a polynomial factor is dominated by any positive exponential, we obtain
\begin{align*}
\int_{u_0\sqrt m<|t|\leq\sqrt{km\log m}}
\mathcal E_{m,\ell}(t)\,dt
=
O\left(e^{-c_{D,9,k}m/2}\right),
\end{align*}
which completes the proof of Proposition 3 by setting \(c_{3,k}=c_{D,9,k}/2\).

\section{Proof of Proposition 4} \label{appendix:proposition4}
\par\noindent{\bfseries Proposition 4.} For every fixed \(k\ge1\),
\begin{align*}
\int_{|t|>\sqrt{km\log m}}
\mathcal E_{m,\ell}(t)\,dt
=O(m^{-1}).
\end{align*}
\par\noindent{\bfseries Proof.} By noting that both characteristic functions in \eqref{eq:smoothed-fourier-bound} have modulus at most one, we obtain
\begin{align*}
\int_{|t|>\sqrt{km\log m}}
\mathcal E_{m,\ell}(t)\,dt
&\le
4\int_{\sqrt{km\log m}}^\infty
\frac{e^{-t^2/(2m)}}{t^2}\,dt\\
&=
\frac4{\sqrt m}
\int_{\sqrt{k\log m}}^\infty
\frac{e^{-u^2/2}}{u^2}\,du\\
&=
O\left(
m^{-(k+1)/2}(\log m)^{-3/2}
\right),
\end{align*}
where the last step follows since 
\begin{align*}
\int_a^\infty\frac{e^{-u^2/2}}{u^2}\,du
\le
\frac{e^{-a^2/2}}{a^3}.
\end{align*}
Therefore, \eqref{eq:smoothing-tail} holds for every \(k\ge1\).  \qed

\section{Proof of Proposition 5} \label{app:desmoothing}

\par\noindent{\bfseries Proposition 5 (Desmoothing error).}
The two desmoothing error terms satisfy
\begin{align*}
\left|
\mathbb E[h(L_{m,\ell}+G_m)]-\mathbb E[h(L_{m,\ell})]
\right|
=O(m^{-1}),
\qquad
\left|
\mathbb E[h(N_{m,\ell}+G_m)]-\mathbb E[h(N_{m,\ell})]
\right|
=O(m^{-1}).
\end{align*}

In Lemma 3 below, we provide an anti-concentration bound that controls the probability that
\(L_{m,\ell}\) lies in any short interval. In particular, it will be
applied to the random interval \([-|G_m|,|G_m|]\) to control the event
on which adding \(G_m\) crosses the origin, where \(h(x) = \exp(-|x|/2)\) is not
differentiable.
\par\noindent{\bfseries Lemma 3 (Uniform anti-concentration bound).}
There exists a constant \(c_{F,1}>0\) such that, uniformly over
\(a\in\mathbb R\) and \(s\geq0\),
\begin{align*}
\Pr\left\{
a\leq L_{m,\ell}\leq a+s
\right\}
\leq
c_{F,1}\left(s+m^{-1/2}\right).
\end{align*}

\par\noindent{\bfseries Proof of Lemma 3.}
For simplicity, we write 
\begin{align*}
\mu_m
\triangleq
\mathbb E_{H_m}\left(  \log\frac{P_m(\mathbf Z_j)}{Q_0^{\otimes m}(\mathbf Z_j)} \right),
\qquad
\sigma_m^2
\triangleq
\operatorname{Var}_{H_m}\left(  \log\frac{P_m(\mathbf Z_j)}{Q_0^{\otimes m}(\mathbf Z_j)} \right).
\end{align*}
In Appendix~\ref{app:midpoint-moments}, we prove that 
\begin{align}
&\mu_m
=
O(m^{-2}),
\quad  
\sigma_{m}^2
=
\frac{\chi_2(Q_1\|Q_0)}{m}
+
O(m^{-2}), \label{eq:moment1}\\
& \mathbb E_{H_m}
\left|
\left(  \log\frac{P_m(\mathbf Z_j)}{Q_0^{\otimes m}(\mathbf Z_j)} \right)-\mathbb E_{H_m} \left(  \log\frac{P_m(\mathbf Z_j)}{Q_0^{\otimes m}(\mathbf Z_j)} \right)
\right|^3
=
O(m^{-3/2}). \label{eq:moment2}
\end{align}
Let \(G_0\sim N(0,1)\), and define the standardized version of $L_{m,\ell}$ as 
\begin{align*}
S_{m,\ell}
\triangleq
\frac{L_{m,\ell}-\ell\mu_m}
{\sqrt{\ell}\sigma_m}.
\end{align*}
By \eqref{eq:moment1} and the assumption \(c_0\leq\ell/m\leq c_1\), we note that 
\begin{align*}
\ell\sigma_m^2
=
\frac{\ell}{m}\chi_2(Q_1\|Q_0)
+
O(m^{-1})
\end{align*}
is bounded above and bounded away from zero. The Berry--Esseen theorem therefore yields
\begin{align*}
\sup_{x\in\mathbb R}
\left|
\Pr\{S_{m,\ell}\leq x\}
-
\Pr\{G_0\leq x\}
\right|
&\leq
c_{F,2}\,
\frac{
\ell\,
\mathbb E_{H_m}\left|\log\frac{P_m(\mathbf Z_j)}{Q_0^{\otimes m}(\mathbf Z_j)}-\mu_m \right|^3
}{
(\ell\sigma_m^2)^{3/2}
} =
O(m^{-1/2}),
\end{align*}
where the last step follows from~\eqref{eq:moment1} and~\eqref{eq:moment2}.
Fix \(a\in\mathbb R\), \(s\geq0\), and \(\eta>0\). Applying
the Berry--Esseen bound twice yields 
\begin{align*}
\Pr\{a\leq L_{m,\ell}\leq a+s\}
&\leq
\Pr\{a-\eta<L_{m,\ell}\leq a+s\} \\
&=
\Pr\left\{
\frac{a-\eta-\ell\mu_m}{\sqrt{\ell}\sigma_m}
<
S_{m,\ell}
\leq
\frac{a+s-\ell\mu_m}{\sqrt{\ell}\sigma_m}
\right\} \\
&\leq
\Pr\left\{
\frac{a-\eta-\ell\mu_m}{\sqrt{\ell}\sigma_m}
<
G_0
\leq
\frac{a+s-\ell\mu_m}{\sqrt{\ell}\sigma_m}
\right\}
+
O(m^{-1/2}).
\end{align*}
The distance between the two endpoints is
$\frac{s+\eta}{\sqrt{\ell}\sigma_m}
\leq
c_{F,3}(s+\eta)$ for some constant $c_{F,3}$,  
since \(\sqrt{\ell}\sigma_m\) is bounded away from zero. Since the density of 
standard Gaussian random variable $G_0$ is bounded by \(1/\sqrt{2\pi}\), we obtain
\begin{align*}
\Pr\{a\leq L_{m,\ell}\leq a+s\}
\leq
\frac{c_{F,3}(s+\eta)}{\sqrt{2\pi}}+O(m^{-1/2}).
\end{align*}
Letting \(\eta\downarrow0\) yields
\begin{align*}
\Pr\{a\leq L_{m,\ell}\leq a+s\}
\leq
\frac{c_{F,3}s}{\sqrt{2\pi}}+O(m^{-1/2})
\leq c_{F,1}\left(s+m^{-1/2}\right).
\end{align*}
The enlargement of the interval at its lower endpoint ensures that
the argument also covers a possible atom at \(a\). \qed

To use the centering of \(G_m\), we need a first-order expansion of
\(h\). The following bound separately accounts for the possibility
that the perturbation crosses the origin, where \(h\) is not
differentiable.

\par\noindent{\bfseries Lemma 4 (Perturbation bound for \(h\)).}
Set \(h'(0)=0\). Then, for all \(x,y\in\mathbb R\),
\begin{align}
\left|
h(x+y)-h(x)-yh'(x)
\right|
\leq
y^2
+
|y|\mathbf 1_{\{|x|\leq|y|\}}.
\label{eq:h-perturbation}
\end{align}

\par\noindent{\bfseries Proof of Lemma 4.}
First, one can verify that for $u \ne 0$,
$|h'(u)|\leq\frac12$ and
$|h''(u)|\leq\frac14.$
Suppose that \(x(x+y)>0\), so that the segment between \(x\)
and \(x+y\) does not contain the origin. Taylor's theorem implies, for
some point \(\xi\) between \(x\) and \(x+y\),
\begin{align*}
h(x+y)
=
h(x)+yh'(x)+\frac{y^2}{2}h''(\xi).
\end{align*}
Therefore, we obtain
\begin{align*}
\left|
h(x+y)-h(x)-yh'(x)
\right|
\leq
\frac{y^2}{2}\sup_{u\neq0}|h''(u)|
\leq
\frac{y^2}{8}.
\end{align*}

It remains to consider \(x(x+y)\leq0\), in which case the segment
between \(x\) and \(x+y\) crosses or touches the origin. This implies
$|x|\leq|y|,$
and hence
$\mathbf 1_{\{|x|\leq|y|\}}=1.$
Moreover, note that \(h\) is \(1/2\)-Lipschitz because
\begin{align*}
|h(v)-h(w)|
\leq
\frac12|v-w|,
\qquad \forall v,w\in\mathbb R.
\end{align*}
Using this property and \(|h'(x)|\leq1/2\), with \(h'(0)=0\), we obtain
\begin{align*}
\left|
h(x+y)-h(x)-yh'(x)
\right|
\leq
|h(x+y)-h(x)|
+
|y|\,|h'(x)|\leq
\frac12|y|+\frac12|y|=
|y|.
\end{align*}
Combining the two cases completes the proof of Lemma~4.  \qed

Applying \eqref{eq:h-perturbation} pointwise with \(x=L_{m,\ell}\) and \(y=G_m\) yields
\begin{align*}
\left|
h(L_{m,\ell}+G_m)
-h(L_{m,\ell})
-G_mh'(L_{m,\ell})
\right| \leq
G_m^2
+
|G_m|
\mathbf 1_{\{|L_{m,\ell}|\leq|G_m|\}}
\end{align*}
almost surely. Since \(G_m \sim N(0,m^{-1}) \) is independent of \(L_{m,\ell}\) and
\(\mathbb E[G_m]=0\), we have 
\begin{align*}
\mathbb E\left[G_mh'(L_{m,\ell})\right]=0.
\end{align*}
Taking expectations in the preceding pointwise inequality and using
\(|\mathbb EX|\leq\mathbb E|X|\), we obtain
\begin{align*}
\left|
\mathbb E[h(L_{m,\ell}+G_m)]
-
\mathbb E[h(L_{m,\ell})]
\right|
\leq
\mathbb E[G_m^2]
+
\mathbb E\left[
|G_m|
\mathbf 1_{\{|L_{m,\ell}|\leq|G_m|\}}
\right].
\end{align*}
Condition on \(G_m=g\). Since \(G_m\) is independent of
\(L_{m,\ell}\), Lemma 3 applied to the interval
\([-|g|,|g|]\), whose length is \(2|g|\), gives
\begin{align*}
\Pr\left\{
|L_{m,\ell}|\leq|G_m|
\,\middle|\,G_m=g
\right\}
=
\Pr\left\{
-|g|\leq L_{m,\ell}\leq|g|
\right\}\leq
2c_{F,1}\left(|g|+m^{-1/2}\right).
\end{align*}
Hence, by the law of iterated expectations,
\begin{align*}
\mathbb E\left[
|G_m|
\mathbf 1_{\{|L_{m,\ell}|\leq|G_m|\}}
\right] &=
\mathbb E\left[
\mathbb E\left[
|G_m|
\mathbf 1_{\{|L_{m,\ell}|\leq|G_m|\}}
\,\middle|\,G_m
\right]
\right] \\
&=
\mathbb E\left[
|G_m|
\Pr\left\{
|L_{m,\ell}|\leq|G_m|
\,\middle|\,G_m
\right\}
\right]\\
&\leq
2c_{F,1}\mathbb E\left[
|G_m|\left(|G_m|+m^{-1/2}\right)
\right]\\
&=
2c_{F,1}\mathbb E G_m^2
+
2c_{F,1}m^{-1/2}\mathbb E|G_m|,
\end{align*}
which scales as $O(m^{-1})$ since the Gaussian variable  \(G_m \sim N(0,m^{-1}) \) satisfies 
$\mathbb E [G_m^2]=\frac1m$ and 
$\mathbb E[|G_m|]=\sqrt{\frac{2}{\pi m}}.$
Therefore,  we obtain
\begin{align*}
\left|
\mathbb E[ h(L_{m,\ell}+G_m)]
-
\mathbb E [h(L_{m,\ell})]
\right|
=
O(m^{-1}).
\end{align*}

Next, we bound the term $\left|
\mathbb E [h(N_{m,\ell}+G_m)]
-
\mathbb E [h(N_{m,\ell})]
\right|$. Since
\begin{align*}
\operatorname{Var}(N_{m,\ell})
=
\frac{\ell\chi_2(Q_1\|Q_0)}{m}
\geq
c_0\chi_2(Q_1\|Q_0)>0,
\end{align*}
the density of \(N_{m,\ell}\) satisfies
\begin{align*}
f_{N_{m,\ell}}(x)
=
\sqrt{
\frac{m}{
2\pi\ell\chi_2(Q_1\|Q_0)
}
}
\exp\left(
-\frac{mx^2}{
2\ell\chi_2(Q_1\|Q_0)
}
\right) \leq
\frac{1}{
\sqrt{
2\pi c_0\chi_2(Q_1\|Q_0)
}
}.
\end{align*}
Conditioning on \(G_m=g\) and using the independence of \(G_m\) and
\(N_{m,\ell}\), we obtain
\begin{align*}
&
\Pr\left\{
|N_{m,\ell}|\leq|G_m|
\,\middle|\,G_m=g
\right\}
=
\Pr\left\{
-|g|\leq N_{m,\ell}\leq|g|
\right\} =
\int_{-|g|}^{|g|}
f_{N_{m,\ell}}(x)\,dx \leq
\frac{2|g|}{
\sqrt{
2\pi c_0\chi_2(Q_1\|Q_0)
}
}.
\end{align*}
Consequently, 
\begin{align*}
\Pr\left\{
|N_{m,\ell}|\leq|G_m|
\,\middle|\,G_m
\right\}
\leq
\frac{2|G_m|}{
\sqrt{
2\pi c_0\chi_2(Q_1\|Q_0)
}
} .
\end{align*}
Applying \eqref{eq:h-perturbation} with \(x=N_{m,\ell}\) and \(y=G_m\) yields, for some constant \(c_{F,4}>0\),
\begin{align*}
&
\left|
\mathbb E [h(N_{m,\ell}+G_m)]
-
\mathbb E [h(N_{m,\ell})]
\right|
 \leq
c_{F,4}\mathbb E[ G_m^2]
+
c_{F,4}\mathbb E\left[
|G_m|
\Pr\left\{
|N_{m,\ell}|\leq|G_m|
\,\middle|\,G_m
\right\}
\right] \leq O(m^{-1}).
\end{align*}

\section{Proof of Proposition 6} \label{appendix:proposition6}
We apply \cite[Theorem 5]{Tahmasbi2019} with the length-\(n\),
weight-\(\ell\) PPM input distribution $\mathcal P_{n,\ell}$. That theorem reduces the
random-coding argument to four verifications: a reliability bound for
each PPM codeword, an information-spectrum bound for channel
resolvability, a positive-probability condition ensuring that the
reliability and resolvability properties hold simultaneously, and a
covertness guarantee for the PPM-induced output distribution. We verify the first
conditions in this order, whereas the last one is exactly the assumption in Proposition~6.

\noindent\emph{1) Choice of the message and key sizes.}
We  choose the message size \(M_n\) such that
\begin{align*}
M_n
\triangleq
\left\lfloor
\exp\left\{
\ell D_P
-\sqrt{\ell V_P}\,
Q^{-1}\left(
\epsilon-\frac{\kappa}{\sqrt{\ell}}
\right)
-13\log n
\right\}
\right\rfloor,
\end{align*}
where \(\kappa>0\) is a sufficiently large 
constant. We choose \(K_n\) as the smallest positive integer
satisfying
\begin{align*}
\log(M_nK_n)
\ge
\max\left\{
\log M_n,\,
\ell D_Q+n^{3/8}
\right\}.
\end{align*}

\medskip
\noindent\emph{2) Verification of the reliability condition.}
Let $P_{Y,\mathrm{PPM}}^{n,\ell}(\mathbf Y)$ be the distribution induced by the PPM input distribution and the channel $W_{Y|X}$. For every PPM codeword $\mathbf{x}$, we define its \emph{information density} as  $\imath(\mathbf x;\mathbf Y)
:=
\sum_{j:x_j=1}
\log\frac{P_1(Y_j)}{P_0(Y_j)}.$
We first verify that, for all sufficiently large \(n\),
\begin{align*}
&\max_{\mathbf x\in\mathcal P_{n,\ell}}
\Pr\left\{
\imath(\mathbf x;\mathbf Y)
\le \log(n^{13}M_n)
\right\} +
\frac1n
\mathbb E_{P_{Y,\mathrm{PPM}}^{n,\ell}}
\left[
\frac{
P_{Y,\mathrm{PPM}}^{n,\ell}(\mathbf Y)
}{
P_0^{\otimes n}(\mathbf Y)
}
\right]
\le\epsilon,
\end{align*}
Applying \cite[Lemma 5]{Tahmasbi2019} yields
\begin{align*}
\Pr\left\{
\imath(\mathbf x;\mathbf Y)
\le \log(n^{13}M_n)
\right\}
\le{}&
Q\left(
\frac{\ell D_P-\log(n^{13}M_n)}
{\sqrt{\ell V_P}}
\right)
+\frac{c_{G,1}}{\sqrt{\ell}},
\end{align*}
where \(c_{G,1}>0\) depends only on the channel.
By the definition of \(M_n\), we have
\begin{align*}
\frac{\ell D_P-\log(n^{13}M_n)}
{\sqrt{\ell V_P}}
=
Q^{-1}\left(
\epsilon-\frac{\kappa}{\sqrt{\ell}}
\right)
+o(\ell^{-1/2}).
\end{align*}
Since \(Q\) is Lipschitz on bounded intervals, we have
\begin{align*}
Q\left(
\frac{\ell D_P-\log(n^{13}M_n)}
{\sqrt{\ell V_P}}
\right)
=
\epsilon-\frac{\kappa}{\sqrt{\ell}}
+o(\ell^{-1/2}).
\end{align*}

The second term in the reliability condition is controlled by
\cite[Lemma 6]{Tahmasbi2019}:
\begin{align*}
\mathbb E_{P_{Y,\mathrm{PPM}}^{n,\ell}}
\left[
\frac{
P_{Y,\mathrm{PPM}}^{n,\ell}(\mathbf Y)
}{
P_0^{\otimes n}(\mathbf Y)
}
\right]
\le
\exp\left(
\frac{\ell(\ell+1)}{n}
\chi_2(P_1\|P_0)
\right).
\end{align*}
Because \(\ell=\Theta(\sqrt n)\), the exponent on the right-hand
side is bounded, and therefore
\begin{align*}
\frac1n
\mathbb E_{P_{Y,\mathrm{PPM}}^{n,\ell}}
\left[
\frac{
P_{Y,\mathrm{PPM}}^{n,\ell}(\mathbf Y)
}{
P_0^{\otimes n}(\mathbf Y)
}
\right]
=O(n^{-1})
=o(\ell^{-1/2}).
\end{align*}
Choosing \(\kappa>c_{G,1}\) sufficiently large gives
\begin{align*}
&\max_{\mathbf x\in\mathcal P_{n,\ell}}
\Pr\left\{
\imath(\mathbf x;\mathbf Y)
\le \log(n^{13}M_n)
\right\}+
\frac1n
\mathbb E_{P_{Y,\mathrm{PPM}}^{n,\ell}}
\left[
\frac{
P_{Y,\mathrm{PPM}}^{n,\ell}(\mathbf Y)
}{
P_0^{\otimes n}(\mathbf Y)
}
\right]
\le\epsilon
\end{align*}
for all sufficiently large \(n\). Thus the reliability condition of
\cite[Theorem 5]{Tahmasbi2019} is satisfied.

\medskip
\noindent\emph{3) Verification of the resolvability condition.}
By the choice of \(K_n\), we have
$\log\frac{M_nK_n}{n^4}
\ge
\ell D_Q+n^{3/8}-4\log n.$
Applying the PPM information-spectrum bound in
\cite[Lemma 7]{Tahmasbi2019}, there exists a channel-dependent
constant \(c_{G,2}>0\) such that
\begin{align*}
&\Pr\left\{
\log
\frac{
W_{Z|X}^{\otimes n}(\mathbf Z|\mathbf X)
}{
P_{Z,\mathrm{PPM}}^{n,\ell}(\mathbf Z)
}
\ge
\log\frac{M_nK_n}{n^4}
\right\} \le
\exp\left[
-c_{G,2}\,
\frac{
\bigl(n^{3/8}-4\log n\bigr)^2
}{
\ell\log^2m
}
\right],
\end{align*}
which is exponentially small since $\ell = \Theta(n^{1/2})$. This verifies the information-spectrum
condition  for resolvability.

\medskip
\noindent\emph{4) Verification of the positive-probability condition.}
The remaining condition in
\cite[Theorem 5]{Tahmasbi2019} guarantees that the reliability and
resolvability properties hold simultaneously for at least one
realization of the random codebook.

Note that $M_n=\exp(\Theta(\sqrt n))$ and $
\log(M_nK_n)=O(\sqrt n)$. The remaining factors in the positive-probability condition  depend on the finite output alphabet,
the minimum positive output probability, and polynomial powers of
\(n\). Since the channel is fixed and the alphabet is finite, the
corresponding logarithmic factors grow at most polynomially in \(n\).
Thus the quantity inside the exponential term in that condition is
bounded below by a positive inverse polynomial in \(n\), whereas
\(M_n=\exp(\Theta(\sqrt n))\). The exponential factor therefore
converges to one, and the positive-probability condition holds for all
sufficiently large \(n\).

We have now verified all the conditions of
\cite[Theorem 5]{Tahmasbi2019}. Consequently, for the covertness
metric considered in the proposition, there exists a deterministic
PPM code, with at
least \(M_n\) messages,  satisfying the required covertness constraint, having
maximum probability of error at most \(\epsilon\).

Finally, we note that 
\begin{align*}
\log M_n
&=
\ell D_P
-\sqrt{\ell V_P}\,
Q^{-1}\left(
\epsilon-\frac{\kappa}{\sqrt{\ell}}
\right)
-13\log n
+o(1)\\
&=
\ell D_P
-\sqrt{\ell V_P}\,Q^{-1}(\epsilon)
+O(\log n),
\end{align*}
since
$Q^{-1}\left(
\epsilon-\frac{\kappa}{\sqrt{\ell}}
\right)
=
Q^{-1}(\epsilon)+O(\ell^{-1/2}),$ and
\begin{align*}
\log K_n
&=
\left[
\ell D_Q+n^{3/8}-\log M_n
\right]^+
+O(1)\\
&=
\left[
\ell(D_Q-D_P)
+n^{3/8}
+O(\sqrt{\ell}+\log n)
\right]^+
+O(1)\\
&=
\ell[D_Q-D_P]^+
+o(\sqrt n).
\end{align*}
This completes the proof of Proposition 6.
\hfill\(\square\)

\section{Estimating the Moment of Log-Likelihood Ratio}
\label{app:midpoint-moments}

This appendix aims to derive  the mean, variance, and third absolute centered
moment of the one-block log-likelihood ratio under the midpoint
distribution $H_m$:
\begin{align}
\mathbb E_{H_m} \left( \log\frac{P_m(\mathbf Z)}{Q_0^{\otimes m}(\mathbf Z)} \right)
=
O(m^{-2}),
\quad
\operatorname{Var}_{H_m}\left(  \log\frac{P_m(\mathbf Z)}{Q_0^{\otimes m}(\mathbf Z)} \right)
=
\frac{\chi_2(Q_1\|Q_0)}{m}
+
O(m^{-2}),
\label{eq:midpoint-mean-variance}
\end{align}
\begin{align}
\mathbb E_{H_m}
\left|
\left(  \log\frac{P_m(\mathbf Z_j)}{Q_0^{\otimes m}(\mathbf Z_j)} \right)-\mathbb E_{H_m} \left(  \log\frac{P_m(\mathbf Z_j)}{Q_0^{\otimes m}(\mathbf Z_j)} \right)
\right|^3
=
O(m^{-3/2}).
\label{eq:midpoint-third-moment}
\end{align}
As a result,  we also have 
\begin{align*}
\operatorname{Var}_{H_m}(X_m) = \chi_2(Q_1 \|Q_0) + O(m^{-1}), \quad \text{and} \quad \mathbb E_{H_m}
\left|
X_m-\mathbb E_{H_m}X_m
\right|^3 = O(1),
\end{align*}
where $X_m$ is defined in~\eqref{eq:xm} in Appendix~\ref{appendix:proposition2}.

\subsection{Mean And Variance}
For simplicity, we abbreviate 
$C_m
\triangleq
\log
\frac{P_m(\mathbf Z)}
{Q_0^{\otimes m}(\mathbf Z)}$, where
$\mathbf Z\sim H_m.$
If \(P_m(\mathbf z)=0\), then \(H_m(\mathbf z)=0\). Therefore, the
value assigned to \(C_m\) at such points does not affect any
expectation under \(H_m\).

Recall that
$M_m(s)
=
\sum_{\mathbf z}
Q_0^{\otimes m}(\mathbf z)
\left(
\frac{P_m(\mathbf z)}
{Q_0^{\otimes m}(\mathbf z)}
\right)^s$ for
$\operatorname{Re}(s)>0.$
Differentiating it with respect to $s$ gives
\begin{align*}
M_m'(s)
&=
\sum_{\mathbf z}
Q_0^{\otimes m}(\mathbf z)
\left(
\frac{P_m(\mathbf z)}
{Q_0^{\otimes m}(\mathbf z)}
\right)^s
\log
\frac{P_m(\mathbf z)}
{Q_0^{\otimes m}(\mathbf z)}.
\end{align*}
At \(s=1/2\),
\begin{align*}
M_m'(1/2)
&=
\sum_{\mathbf z}
Q_0^{\otimes m}(\mathbf z)
\sqrt{
\frac{P_m(\mathbf z)}
{Q_0^{\otimes m}(\mathbf z)}
}
\log
\frac{P_m(\mathbf z)}
{Q_0^{\otimes m}(\mathbf z)}\\
&=
\sum_{\mathbf z}
\sqrt{
P_m(\mathbf z)Q_0^{\otimes m}(\mathbf z)
}
\log
\frac{P_m(\mathbf z)}
{Q_0^{\otimes m}(\mathbf z)} = \rho_m\mathbb E_{H_m}[C_m],
\end{align*}
Also note that \(M_m(1/2)=\rho_m\). Therefore, we have
\begin{align*}
\left.
\frac{d}{ds}\log M_m(s)
\right|_{s=1/2}
=
\frac{M_m'(1/2)}{M_m(1/2)}
=
\mathbb E_{H_m}[C_m].
\end{align*}

Differentiating \(M_m(s)\) with respect to $s$ twice gives
\begin{align*}
M_m''(s)
&=
\sum_{\mathbf z}
Q_0^{\otimes m}(\mathbf z)
\left(
\frac{P_m(\mathbf z)}
{Q_0^{\otimes m}(\mathbf z)}
\right)^s
\left(
\log
\frac{P_m(\mathbf z)}
{Q_0^{\otimes m}(\mathbf z)}
\right)^2.
\end{align*}
Evaluating at \(s=1/2\) and again using the definition of \(H_m\),
we obtain
\begin{align*}
M_m''(1/2)
=
\rho_m\mathbb E_{H_m}[C_m^2].
\end{align*}
Consequently, we have
\begin{align*}
\left.
\frac{d^2}{ds^2}\log M_m(s)
\right|_{s=1/2}
&=
\frac{M_m''(1/2)}{M_m(1/2)}
-
\left(
\frac{M_m'(1/2)}{M_m(1/2)}
\right)^2 =
\mathbb E_{H_m}[C_m^2]
-
\bigl(\mathbb E_{H_m}C_m\bigr)^2 =
\operatorname{Var}_{H_m}(C_m).
\end{align*}

It remains to evaluate these derivatives using Lemma~2:
\begin{align*}
\log M_m(s)
=
\frac{\chi_2(Q_1\|Q_0)}{2m}s(s-1)
+
r_m(s), \quad \text{where }  |r_m(s)|
\le
\frac{c_{B,1,d}(1+|s|)^4}{m^2}.
\end{align*}
The derivative of the first term is
\begin{align*}
\frac{d}{ds}
\left[
\frac{\chi_2(Q_1\|Q_0)}{2m}s(s-1)
\right]
=
\frac{\chi_2(Q_1\|Q_0)}{2m}(2s-1),
\end{align*}
which equals zero when \(s=1/2\). Its second derivative is
\begin{align*}
\frac{d^2}{ds^2}
\left[
\frac{\chi_2(Q_1\|Q_0)}{2m}s(s-1)
\right]
=
\frac{\chi_2(Q_1\|Q_0)}{m}.
\end{align*}

We next  differentiate the remainder term \(r_m(s)\). Choose the
fixed disk
$\mathcal D
\triangleq
\left\{
s\in\mathbb C:
\left|s-\frac12\right|\le\frac18
\right\}.$
Since \(|s|\) is bounded on
\(\mathcal D\), we have 
$\sup_{s\in\mathcal D}|r_m(s)|
\le
\frac{c_{H,1}}{m^2}$ for some constant $c_{H,1}>0$.
The function \(r_m(s)\) is analytic on this disk. Cauchy's derivative
estimate therefore gives, for \(k=1,2\),
\begin{align*}
\left|
r_m^{(k)}\left(\frac12\right)
\right|
\le
\frac{k!}{(1/8)^k}
\sup_{s \in \mathcal{D}}
|r_m(s)|
=
O(m^{-2}).
\end{align*}
It follows that
\begin{align*}
\mathbb E_{H_m}[C_m]
&=
\left.
\frac{d}{ds}\log M_m(s)
\right|_{s=1/2} =
r_m'(1/2) =
O(m^{-2}),
\end{align*}
\begin{align*}
\operatorname{Var}_{H_m}(C_m)
&=
\left.
\frac{d^2}{ds^2}\log M_m(s)
\right|_{s=1/2} =
\frac{\chi_2(Q_1\|Q_0)}{m}
+
r_m''(1/2) =
\frac{\chi_2(Q_1\|Q_0)}{m}
+
O(m^{-2}).
\end{align*}
This completes the proof of \eqref{eq:midpoint-mean-variance}.

\subsection{Third Absolute Centered Moment}

The proof uses the event \(\mathcal E_m\) introduced in Section~\ref{sec:variational} and its complement \(\mathcal E_m^{\mathrm c}\).
We first control the contribution from \(\mathcal E_m\). Note that 
\begin{align*}
\mathbb E_{H_m}
\left[
|C_m|^3\mathbf 1_{\mathcal E_m}
\right]
&=
\sum_{\mathbf z\in\mathcal E_m}
H_m(\mathbf z)|C_m(\mathbf z)|^3
\end{align*}
For
\(|x|\le1/2\),
we have $|\log(1+x)| \le
2|x|.$
Therefore, on the event \(\mathcal E_m\), we have 
$|C_m|
=
|\log(1+B_m)|
\le
2|B_m|.$
The definition of \(H_m\) gives
\begin{align*}
H_m(\mathbf z)
=
Q_0^{\otimes m}(\mathbf z)
\frac{\sqrt{1+B_m(\mathbf z)}}{\rho_m}.
\end{align*}
On the event \(\mathcal E_m\), we have
$\sqrt{1+B_m(\mathbf z)}
\le
\sqrt{\frac32}.$
Moreover, since \(\rho_m\to1\), we have \(\rho_m\ge1/2\) for all sufficiently
large \(m\). Hence, on the event \(\mathcal E_m\),
$H_m(\mathbf z)
\le
2\sqrt{\frac32}\,
Q_0^{\otimes m}(\mathbf z).$
Therefore,
\begin{align*}
\mathbb E_{H_m}
\left[
|C_m|^3\mathbf 1_{\mathcal E_m}
\right]
&\le
2\sqrt{\frac32}
\sum_{\mathbf z\in\mathcal E_m}
Q_0^{\otimes m}(\mathbf z)
\bigl(2|B_m(\mathbf z)|\bigr)^3 \le
16\sqrt{\frac32}\,
\mathbb E_{Q_0^{\otimes m}}|B_m|^3 = O(m^{-3/2}),
\end{align*}
where the last step follows from the fact that $\mathbb E |B_m|^3 =O(m^{-3/2})$ (calculated in Appendix~\ref{app:lemma2}).

We next control the complement event \(\mathcal E_m^{\mathrm c}\). 
Note that  for all sufficiently large \(m\),
\begin{align*}
H_m(\mathbf z)
&=
Q_0^{\otimes m}(\mathbf z)
\frac{\sqrt{1+B_m(\mathbf z)}}{\rho_m} \le
2\sqrt{a_{\max}+1}\,
Q_0^{\otimes m}(\mathbf z).
\end{align*}
Consequently,
\begin{align*}
\mathbb E_{H_m}
\left[
|C_m|^3\mathbf 1_{\mathcal E_m^{\mathrm c}}
\right]
&=
\sum_{\mathbf z\in\mathcal E_m^{\mathrm c}}
H_m(\mathbf z)|C_m(\mathbf z)|^3 \le
2\sqrt{a_{\max}+1} \times \max_{\mathbf{z}: H_m(\mathbf{z}) > 0 } |\log(1+ B_m(\mathbf{z}))|^3 \times 
Q_0^{\otimes m}(\mathcal E_m^{\mathrm c}).
\end{align*}
Using the concentration inequality in \eqref{eq:Bm-tail} and the fact that $|\max_{\mathbf{z}: H_m(\mathbf{z}) > 0 }\log(1+ B_m(\mathbf{z}))| = O( \log m)$, we obtain 
\begin{align*}
\mathbb E_{H_m}
\left[
|C_m|^3\mathbf 1_{\mathcal E_m^{\mathrm c}}
\right]
= O(e^{-(1/(2(a_{\max}-a_{\min})^2))m}).
\end{align*}
Combining the two events together yields that
\begin{align*}
\mathbb E_{H_m}|C_m|^3
=
O(m^{-3/2}).
\end{align*}
Finally, one can check that 
\begin{align*}
\mathbb E_{H_m}
\left|
C_m-\mathbb E_{H_m}C_m
\right|^3
&\le
4\mathbb E_{H_m}|C_m|^3
+
4\left|\mathbb E_{H_m}C_m\right|^3 = O(m^{-3/2}).
\end{align*}
This completes the proof of \eqref{eq:midpoint-third-moment}.

\end{document}